\documentclass[final,3p,times,11pt]{elsarticle}

\usepackage{amsmath,amssymb,amsthm}
\usepackage{cases}
\usepackage{algorithm}
\usepackage{algpseudocode}
\usepackage{sidecap}
\usepackage{graphicx}
\usepackage{datetime}
\usepackage{bm}

\newcommand{\comment}[1]{}

\journal{Theoretical Computer Science}

\newcommand{\R}{\mathbb{R}}               
\newcommand{\N}{\mathbb{N}}               
\newcommand{\bH}{\mathbf{H}}              
\newcommand{\bL}{\mathbf{L}}              
\newcommand{\Prn}{\Pr\nolimits}           
\DeclareMathOperator{\Exp}{\mathbf{E}}    
\DeclareMathOperator{\Var}{\mathbf{Var}}  
\newcommand{\LAs}{\underline{N}}          
\newcommand{\NAs}{\underline{M}}          
\DeclareMathOperator{\Regret}{R}          
\DeclareMathOperator{\wcRegret}{\bar{\Regret}} 
\DeclareMathOperator{\rank}{rank}         
\algnewcommand\algorithmicto{\textbf{to}}
\newcommand{\one}[1]{\mathbb{I}\left(#1\right)} 
\newcommand{\ra}{\Rightarrow}
\newcommand{\B}{\mathcal{B}}
\newcommand{\C}{\mathcal{C}}
\newcommand{\D}{\mathcal{D}}              
\newcommand{\F}{\mathcal{F}}
\renewcommand{\H}{\mathcal{H}}            
\newcommand{\I}{\mathcal{I}}
\newcommand{\OO}{\mathcal{O}}             
\newcommand{\RR}{\mathcal{R}}             
\newcommand{\hr}{\hat{\rho}}              
\newcommand{\hRegret}{\widehat{\Regret}}
\DeclareMathOperator{\Int}{int}
\DeclareMathOperator{\rInt}{rint}         
\newcommand{\Smax}{{S_{\!\rm max}}}
\newcommand{\lamax}{{\lambda_{\rm max}}}
\newcommand{\lamin}{{\lambda_{\rm min}}}
\newcommand{\Nr}{N_{\rm rev}}
\newcommand{\tNr}{{\tilde{N}_{\rm rev}}}
\newcommand{\nature}{the opponent}
\newcommand{\allrev}{all-revealing}
\newcommand{\partrev}{partially-revealing}
\newcommand{\nonrev}{non-revealing}
\newcommand{\ie}{{i.e.},\ }
\newcommand{\eg}{{e.g.},\ }

\newcommand{\ep}{\bm{\varepsilon}}
\newcommand{\Lebesgue}{\lambda}

\newtheorem*{sep-condition}{Separation Condition}
\newtheorem*{non-deg-condition}{Non-de\-gen\-er\-a\-cy Condition}
\newtheorem*{bernstein}{Bernstein's inequality for martingales}
\newtheorem{lemma}{Lemma}
\newtheorem{theorem}[lemma]{Theorem}	
\newdefinition{definition}[lemma]{Definition}
\newdefinition{example}[lemma]{Example}

\newcommand{\AppleTree}{\textsc{AppleTree}}
\newcommand{\Algfig}{Figure}
\newcommand{\Algfigs}{Figures}
\newcommand{\twocolumns}[6]{
	\begin{minipage}[#3]{#1\textwidth}
		#5
	\end{minipage}
	\hspace{0.2cm}
	\begin{minipage}[#4]{#2\textwidth}
		#6
	\end{minipage}
}

\begin{document}

\begin{frontmatter}

\title{Toward a Classification of Finite Partial-Monitoring Games\tnoteref{title1}}

\author{Andr\'as Antos}				\ead{antos@cs.bme.hu} \ead[URL]{http://www.cs.bme.hu/~antos}

\address{Machine Learning Group, Computer and Automation Research Institute of the Hungarian Academy of Sciences, \\ 13-17 Kende utca, H-1111 Budapest, Hungary}

\author{G\'abor Bart\'ok\corref{corresponding1}} \ead{bartok@cs.ualberta.ca} \ead[URL]{http://www.ualberta.ca/~bartok}
\author{D\'avid P\'al}                           \ead{dpal@cs.ualberta.ca} \ead[URL]{http://www.ualberta.ca/~dpal}
\author{Csaba Szepesv\'ari}                      \ead{szepesva@cs.ualberta.ca} \ead[URL]{http://www.ualberta.ca/~szepesva}

\address{Department of Computing Science, University of Alberta, Edmonton, Alberta, T6G 2E8, Canada}

\cortext[corresponding1]{Corresponding authors}
\tnotetext[title1]{Preliminary version of this paper appeared at ALT 2010, September 6--8, 2010, Canberra, Australia~\cite{Bartok-Pal-Szepesvari-2010}. This work was supported in part by AICML, AITF (formerly iCore and AIF), NSERC and the PASCAL2
Network of Excellence under EC grant no. 216886.}

\begin{abstract}
Partial-monitoring games constitute
 a mathematical framework for sequential decision making problems with imperfect feedback:
The learner repeatedly chooses an action,
 \nature\ responds with an outcome,
 and then the learner suffers a loss and receives a feedback signal,
 both of which are fixed functions of the action and the outcome.
The goal of the learner is to minimize his total cumulative loss.
We make progress towards the classification of these games based on their minimax expected regret.
Namely, we classify almost all games with two outcomes and a finite number of actions:
We show that their minimax expected regret is either
 zero, $\widetilde{\Theta}(\sqrt{T})$, $\Theta(T^{2/3})$, or $\Theta(T)$,
 and we give a simple and efficiently computable classification of these four classes of games.
Our hope is that the result can serve
 as a stepping stone toward classifying all finite partial-monitoring games.
\end{abstract}


\end{frontmatter}

\section{Introduction}
\label{section:introduction}
Partial-monitoring games constitute a mathematical framework for sequential decision making problems with imperfect feedback.
They arise as a natural generalization of many sequential decision making problems
 with full or partial feedback such as
 learning with expert advice~\citep{Littlestone-Warmuth-1994,Freund-Schapire-1997,Cesa-Bianchi-Freund-Haussler-Helmbold-Schapire-Warmuth-1997},
 the multi-armed bandit problem~\cite{Auer-Cesa-Bianchi-Freund-Schapire-2002,
 Bubeck-Munos-Stoltz-Szepesvari-2008,Kleinberg-Slivkins-Upfal-2008},
 label efficient prediction~\cite{Helmbold-Panizza-1997,Cesa-Bianchi-Lugosi-Stoltz-2005},
 dynamic pricing~\cite{Kleinberg-Leighton-2003, Blum-Hartline-2005},
 the dark pool problem~\cite{Agarwal-Bartlett-Dama-2010},
 the apple tasting problem~\cite{Helmbold-Littlestone-Long-2000},
 online convex optimization~\cite{Zinkevich-2003a, Zinkevich-2003b},
 online linear~\cite{Abernethy-Hazan-Rakhlin-2008} and
 convex optimization with bandit feedback~\cite{Flaxman-Kalai-McMahan-2005}.

A partial-monitoring game is a repeated game between two players:
 the \emph{learner} and the \emph{opponent}.
In each round, the learner chooses an action and simultaneously \nature\ chooses an outcome.
Next, the learner receives a feedback signal and suffers a loss;
 however neither the loss nor the outcome are revealed to the learner.
The feedback and the loss are fixed functions of the action and the outcome,
 and these functions are known by both players.
The main feature of this model is that
 it captures that the learner has imperfect or partial information about the outcome sequence.
In this work, we make the natural assumption that the opponent is \emph{oblivious},
 that is, the opponent does not have access to the learner's actions.

The goal of the learner is to keep his cumulative loss small.
However, since \nature\ could choose the outcome sequence
 so that the learner suffers as high loss as possible,
 it is too much to ask for an absolute guarantee for the cumulative loss.
Instead, a competitive viewpoint is taken and
 the cumulative loss of the learner is compared with the cumulative loss of the best among all the constant strategies,
 \ie strategies that choose the same action in every round.
The difference between the cumulative loss of the learner and the cumulative loss of the best constant strategy
 is called the \emph{regret}.

Generally, the regret grows with the number of rounds of the game.
If the growth is sublinear then the learner is said to be Hannan consistent%
\footnote{Hannan consistency is named after James Hannan who was the first to design a learning algorithm with sublinear regret for finite games with full feedback~\cite{Hannan-1957}.},
 and in the long run the learner's average loss per round approaches the average loss per round of the best action.

Designing learning algorithms with low regret is the main focus of study of partial-monitoring games.
For a given game, the ultimate goal is to find out
 its optimal worst-case (minimax) regret, and design an algorithm that achieves it.
The minimax regret can be viewed as an inherent measure of how hard the game is for the learner.
The motivation behind this paper was the
 desire to determine the minimax regret and design an algorithm achieving it for each game in a large class.

In this paper we restrict our attention
 to games with a finite number of actions and \emph{two outcomes}.
This class is a subset of the class of \emph{finite partial-monitoring games},
 introduced by Piccolboni and Schindelhauer~\cite{Piccolboni-Schindelhauer-2001},
 in which both the set of actions and the set of outcomes are finite.

\subsection{Previous Results}
For full-information games (\ie when the feedback determines the outcome) with $N$
 actions and losses lying in the interval $[0,1]$, there exists a randomized
 algorithm with expected regret at most $\sqrt{T \ln(N)/2}$ where $T$ is the time horizon
 (see \eg \citet[Chapter 4]{Cesa-Bianchi-Lugosi-2006} and references therein).
Furthermore, it is known that this upper bound is tight:
There exist full-information games with losses lying in the interval $[0,1]$
 for which the worst-case expected regret of any algorithm
 is at least $\Omega(\sqrt{T\ln N})$~\cite[Chapter 3]{Cesa-Bianchi-Lugosi-2006}.

Another special case of partial-monitoring games is the multi-armed bandit game,
 where the learner's feedback is the loss of the action he chooses.
For a multi-armed bandit game with $N$ actions and losses lying in the interval $[0,1]$,
 the INF algorithm~\cite{Audibert-Bubeck-2009} has expected regret at most $O(\sqrt{TN})$.
(The well-known Exp3 algorithm~\cite{Auer-Cesa-Bianchi-Freund-Schapire-2002} achieves the bound $O(\sqrt{TN \log N})$.)
It is also known that the bound $O(\sqrt{TN})$ is optimal~\cite{Auer-Cesa-Bianchi-Freund-Schapire-2002}.

Piccolboni and Schindelhauer~\cite{Piccolboni-Schindelhauer-2001} introduced
 finite partial-monitoring games.
They showed that, for any finite game, either
 there is a strategy for the learner that achieves regret of at
 most $O(T^{3/4}(\ln T)^{1/2})$ or the worst-case expected regret of any learner is $\Omega(T)$.
\citet{Cesa-Bianchi-Lugosi-Stoltz-2006}
 improved this result and showed that \citeauthor{Piccolboni-Schindelhauer-2001}'s algorithm achieves $O(T^{2/3})$ regret.
They also gave an example of a game with worst-case expected regret at least $\Omega(T^{2/3})$.
More recently, \citet{Lugosi-Mannor-Stoltz-2008} designed algorithms and proved upper bounds in a slightly different setting,
 where the feedback signal is a possibly noisy function of the outcome or both the action and the outcome.

However, from these results it is unclear what determines which games have
 minimax regret $\Theta(\sqrt{T})$, which games have minimax regret
 $\Theta(T^{2/3})$ and whether there exist finite games with minimax regret not
 belonging to either of these categories.
Cesa-Bianchi et al.~\cite{Cesa-Bianchi-Lugosi-Stoltz-2006} note that:
``\emph{It remains a challenging problem to characterize the class of problems
 that admit rates of convergence faster than $O(n^{-1/3})$.}''%
\footnote{They used $n$ instead of $T$ and by rate they mean the average regret per time step.}

\subsection{Our Results}
We classify the minimax expected regret of finite partial-monitoring games with \emph{two outcomes}.
From our classification we exclude certain ``degenerate games'';
 their precise definition is given later in the paper.
We show that
 the minimax regret of any non-degenerate game falls into one of the four categories:
 $0$, $\widetilde \Theta(\sqrt{T})$, $\Theta(T^{2/3})$, $\Theta(T)$
 and no other option is possible%
\footnote{The notation $\widetilde \Theta$ and $\widetilde O$ hides poly-logarithmic factors in $T$.}.
We call the four classes of games
 \emph{trivial}, \emph{easy}, \emph{hard}, and \emph{hopeless}, respectively.
We give a simple and efficiently computable geometric characterization of these four classes.

Additionally, we show that each of the four classes admits a computationally
 efficient learning algorithm achieving the minimax
 expected regret, up to logarithmic factors.
In particular, we design an efficient learning algorithm for easy games with expected regret at most $\widetilde O(\sqrt{T})$.
For hard games, the algorithm of Cesa-Bianchi et al.~\cite{Cesa-Bianchi-Lugosi-Stoltz-2006} has $O(T^{2/3})$ regret.
For trivial games,
 a simple algorithm that chooses the same action in every round has zero regret.
For hopeless games, any algorithm has $\Theta(T)$ regret.

\section{Basic Definitions and Notations}
\label{section:definitions}
A finite partial-monitoring game is specified by a pair of $N \times M$ matrices $(\bL,\bH)$
 where $N$ is the number of actions, $M$ is the number of outcomes,
 $\bL$ is the \emph{loss matrix}, and $\bH$ is the \emph{feedback matrix}.
We use the notation $\underline{n}=\{1,\dots,n\}$ for any integer
 and denote the actions and outcomes by integers starting from $1$,
 so the action set is $\LAs$ and the outcome set is $\NAs$.
We denote by $\ell_{i,j}$ and $h_{i,j}$ ($i\in\LAs$, $j\in\NAs$) the entries of $\bL$ and $\bH$, respectively.
We denote by $\ell_i$ the $i$-th row ($i\in\LAs$) of $\bL$,
 and we call it the \emph{loss vector of action $i$}.
The elements of $\bL$ are arbitrary real numbers.
The elements of $\bH$ belong to some alphabet $\Sigma$,
 we only assume that the learner is able to distinguish two different elements of the alphabet.
We often use the set of natural or real numbers as the alphabet.

The matrices $\bL$, $\bH$ are known by both the learner and \nature.
The game proceeds in $T$ rounds.
In each round $t=1,2,\dots,T$,
 the learner chooses an action $I_t\in\LAs$
 and simultaneously \nature\ chooses an outcome $J_t\in\NAs$,
 then the learner receives the feedback $h_{I_t,J_t}$.
Nothing else is revealed to the learner;
 in particular $J_t$ and the loss $\ell_{I_t,J_t}$ remain hidden.

In principle, both $I_t$ and $J_t$ can be chosen randomly.
However, to simplify our treatment,
 we assume that \nature\ is deterministic and oblivious to the actions of the learner.
Equivalently, we can assume that the
 sequence of outcomes $J_1, J_2, \dots, J_T$ is a fixed deterministic sequence
 chosen before the first round of the game.
On the other hand, it is important to allow the learner to choose his actions $I_t$ randomly.
A randomized strategy (algorithm) $A$ of the learner is a sequence of random functions $I_1, I_2, \dots, I_T$
 where each of the functions
 maps the feedback from the past outcomes (and learner's internal random ``bits'') to an action;
 formally $I_t:\Sigma^{t-1} \times \Omega \to \LAs$.

The learner is scored according to the loss matrix.
In each round $t$, the learner incurs \emph{instantaneous loss} $\ell_{I_t,J_t}$.
The goal of the learner
 is to keep his \emph{cumulative loss} $\sum_{t=1}^T \ell_{I_t,J_t}$ small.
The \emph{(cumulative) regret} of an algorithm $A$ is defined as
$$
 \hRegret_T = \hRegret_T(A,G)
 = \sum_{t=1}^T \ell_{I_t,J_t} - \min_{i\in\LAs} \sum_{t=1}^T \ell_{i,J_t} \;.
$$
In other words, the regret is the excess loss of the learner
 compared to the loss of the best constant action.
We denote by $\Regret_T = \Regret_T(A,G) = \Exp[\hRegret_T(A,G)]$
 the \emph{(cumulative) expected regret}.
Let the \emph{worst-case expected regret} of $A$ when used in $G=(\bL,\bH)$ be
$$
 \wcRegret_T(A,G) = \sup_{J_{1:T} \in \NAs^T} \ \Regret_T(A,G)\;,
$$
where the supremum is taken over all outcome sequences $J_{1:T} = (J_1, J_2, \dots, J_T) \in \NAs^T$.
The \emph{minimax expected regret} of $G$ (or \emph{minimax regret}, for short) is:
$$
 \Regret_T(G) = \inf_A \ \wcRegret_T(A,G)
 = \inf_A \ \sup_{J_{1:T} \in \NAs^T} \ \Regret_T(A,G)\;,
$$
where the infimum is taken over all randomized strategies $A$.
Note that, since $\Regret_T(A,G)\ge 0$ for constant outcome sequences,
 $\Regret_T(G)\ge 0$ also holds.

We identify the set of all probability distributions over the set of outcomes $\NAs$
 with the 
 probability simplex $\Delta_M = \{ p \in \R^M ~:~ \sum_{j=1}^M p(j) = 1, \ \forall j\in\NAs, \ \ p(j) \ge 0 \}$.
We use $\langle \cdot, \cdot \rangle$ to denote the standard dot product.

\section{Characterization of Games with Two Outcomes}
\label{section:characterization}
In this section, we formally phrase our main characterization result.
We need a preliminary definition that is useful for any finite game:
\begin{definition}[Properties of Actions]
\label{def:action_prop}
Let $G=(\bL,\bH)$ be a finite partial-monitoring game
 with $N$ actions and $M$ outcomes.
Let $i\in\LAs$ be one of its actions.
\begin{itemize}
\item Action $i$ is called \emph{dominated} if for any $p \in \Delta_M$ there exists an action $i'$
 such that $\ell_{i'}\neq\ell_i$ and $\langle \ell_{i'}, p \rangle \le \langle \ell_i, p \rangle$.
\item Action $i$ is called \emph{non-dominated} if it is not dominated.
\item Action $i$ is called \emph{degenerate} if it is dominated and
 there exists a distribution $p\in\Delta_M$ such that
 for all $i'\in\LAs$, $\langle\ell_{i},p\rangle \leq \langle\ell_{i'},p\rangle$.
\item Action $i$ is called \emph{\allrev} if any pair of outcomes $j,j'$, $j \neq j'$ satisfies $h_{i,j} \neq h_{i,j'}$.
\item Action $i$ is called \emph{none-revealing} if any pair of outcomes $j$,$j'$ satisfies $h_{i,j} = h_{i,j'}$.
\item Action $i$ is called \emph{\partrev} if it is neither \allrev\ nor none-revealing. 
\item All-revealing and \partrev\ actions together are called \emph{revealing} actions.
\item Two or more actions with the same loss vector are called \emph{duplicate} actions.
\end{itemize}
\end{definition}
The property of being dominated has an equivalent dual definition.
Namely, action $i$ is dominated if there exists a set of actions with loss vectors not equal to $\ell_i$
 such that some convex combination of their loss vectors is componentwise upper bounded by $\ell_i$.

In games with $M=2$ outcomes, each action is either \allrev\ or none-revealing.
This dichotomy is one of the key properties that lead to the classification theorem for two-outcome games.
To emphasize the dichotomy, from now on we will refer to them as revealing and \emph{\nonrev} whenever it is clear from the context that $M=2$.

The above property also allows us to assume without of loss generality that
 there are no duplicate actions.
Clearly, if multiple actions with the same loss vector exist,
 all but one can be removed (together with the corresponding rows of $\bL$ and $\bH$) without changing the minimax regret:
If all of them are \nonrev, we keep one of the actions and remove all the others.
Otherwise, we keep a revealing action and remove the others.
Then replacing any algorithm by one that, instead of a removed action,
 chooses always the corresponding kept action,
 its loss cannot increase and equals to the loss of this algorithm for the original game.
So the two games have the same minimax regret.

\begin{figure}
\centering
\includegraphics{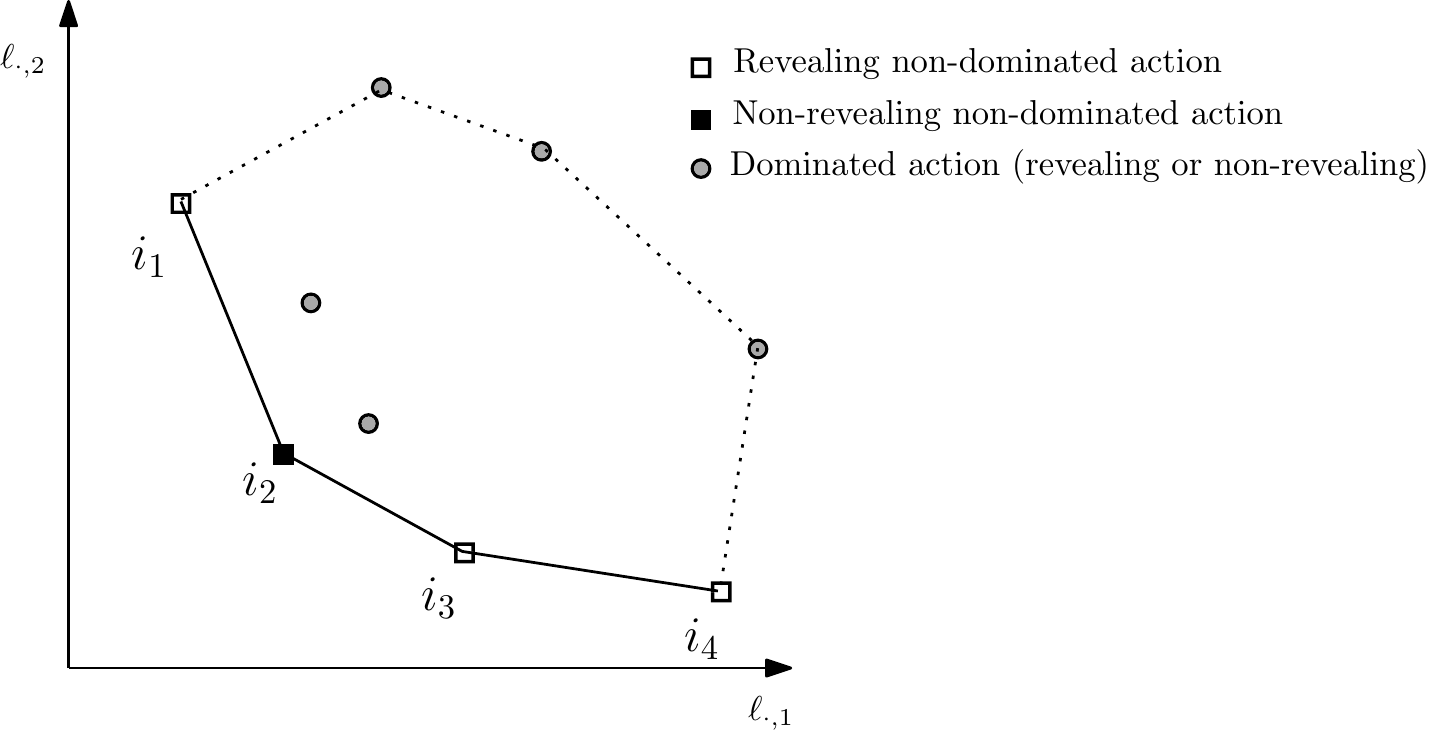}
\caption{
The figure shows each action $i$ as a point in $\R^2$ with coordinates $(\ell_{i,1}, \ell_{i,2})$.
The solid line connects the chain of non-dominated actions,
 which, by convention are ordered according to their loss for the first outcome.
}
\label{fig:convex-chain1}
\end{figure}

The concepts of dominated and non-dominated actions can be visualized for two-outcome games
 by drawing the loss vector of each action as a point in $\R^2$.
The points corresponding to the non-dominated actions
 lie on the bottom-left boundary of the convex hull of the set of all the actions,
 as shown in Figure~\ref{fig:convex-chain1}.
Enumerating the non-dominated actions ordered according to their loss for the first outcome
 gives rise to a sequence $(i_1, i_2, \dots, i_K)$,
 which we call the \emph{chain of non-dominated actions}.

To state the classification theorem, we introduce the following conditions.
\begin{sep-condition} A two-outcome game $G$ satisfies the \emph{separation condition} if,
 after removing duplicate actions,
 its chain of non-dominated actions does \textbf{not} have a pair of consecutive actions $i_k$, $i_{k+1}$
 such that both of them are \nonrev.
The set of games satisfying this condition will be denoted by $\mathcal{S}$.
\end{sep-condition}
\begin{non-deg-condition}
A two-outcome game $G$ is \emph{degenerate}
 if it has a degenerate revealing action.
If $G$ is not degenerate, we call it \emph{non-degenerate}
 and we say that it satisfies the \emph{non-degeneracy condition}.
\end{non-deg-condition}

As we will soon see,
 the separation condition is the key to distinguish between \emph{hard} and \emph{easy} games.
On the other hand,
 the non-de\-gen\-er\-a\-cy condition is merely a technical condition that we need in our proofs.
The set of degenerate games is excluded from the characterization,
 as we do not know the minimax regret of these games.
We are now ready to state our main result.
\begin{theorem}[Classification of Two-Outcome Partial-Monitoring Games]
\label{theorem:characterization}
Let $\mathcal{S}$ be the set of all finite partial-monitoring games with two outcomes
 that satisfy the separation condition.
Let $G=(\bL,\bH)$ be a game with two outcomes that satisfies the non-de\-gen\-er\-a\-cy condition.
Let $K$ be the number of non-dominated actions in $G$, counting duplicate actions only once.
The minimax expected regret $\Regret_T(G)$ satisfies
\begin{subnumcases}{\Regret_T(G)=}
\label{eq:char1} 0 \quad(\forall T), & $K=1$;\\
\label{eq:char2} \widetilde{\Theta}\left(\sqrt{T}\right), & $K\geq2$, $G\in\mathcal{S}$;\\
\label{eq:char3} \Theta\left(T^{2/3}\right), & $K\geq2$, $G\not\in\mathcal{S}$, $G$ has a revealing action;\\
\label{eq:char4} \Theta(T), & otherwise.
\end{subnumcases}
\end{theorem}

We call the games in cases \eqref{eq:char1}--\eqref{eq:char4}
 \emph{trivial}, \emph{easy}, \emph{hard}, and \emph{hopeless}, respectively.
Case~\eqref{eq:char1} is proven by the following lemma
 which shows that a trivial game is also characterized by having $0$ minimax regret in a single round
 or by having an action ``dominating'' alone all the others:
%
\begin{lemma}\label{lem:0minimaxregret}
For any finite partial-monitoring game,
 the following four statements are equivalent:
\begin{enumerate}[a)]
\item The minimax regret is zero for each $T$.
\item The minimax regret is zero for some $T$.
\item There exists a (non-dominated) action $i\in\LAs$
 whose loss is not larger than the loss of any other action
 irrespectively of the choice of Nature's action.
\item The game is trivial, \ie $K=1$ (using the definition in Theorem~\ref{theorem:characterization}).
\end{enumerate}
\end{lemma}
The proof of this lemma can be found in the Appendix.
Case~\eqref{eq:char4} of Theorem~\ref{theorem:characterization} is proven in the Appendix as well.
The upper bound of case~\eqref{eq:char3} can be derived from a result of~\citet{Cesa-Bianchi-Lugosi-Stoltz-2006}:
Recall that the entries of $\bH$ can
 be changed without changing the information revealed to the learner as long as
 one does not change the pattern of which elements in a row are equal and different.
\citet{Cesa-Bianchi-Lugosi-Stoltz-2006} show that if the entries of
 $\bH$ can be chosen such that
 $\rank(\bH)=\rank\left( \begin{array}{c} \bH \\ \bL \\ \end{array} \right)$
 then $O(T^{2/3})$ expected regret is achievable.
This condition holds trivially for two-outcome games with at least one revealing action and $N\ge 2$.
It remains to prove
 the upper bound for case~\eqref{eq:char2},
 the lower bound for~\eqref{eq:char2},
 and the lower bound for~\eqref{eq:char3};
 we prove these in Sections~\ref{section:upper-bound},
 \ref{section:lower-bound-easy-games},
 and \ref{section:lower-bound-hard-games}, respectively.

\section{Examples}
\label{section:examples}

Before we dive into the proof of Theorem~\ref{theorem:characterization},
 we give a few examples of finite partial-monitoring games with two outcomes and
 show how the theorem can be applied.
For each example we present the matrices $\bL,\bH$
 and depict the loss vectors of actions as points in $\R^2$.
\begin{example}[One-Armed Bandit]\label{ex:one-armed}
We start with an example of a multi-armed bandit game.
Multi-armed bandit games are those where the feedback equals
 the instantaneous loss, that is, when $\bL=\bH$.
\footnote{``Classically'', non-stochastic multi-armed bandit problems are defined by the restriction that
 in no round Learner can gain any information about the losses of actions other than the chosen one,
 that is, $\bL$ is not known in advance to Learner.
(Also, the domain set of losses is often infinite there ($M=\infty$).)
When $\bH=\bL$ in our setting, depending on $\bL$, this might or might not be the case;
 the ``classical bandit'' problem with losses constrained to a finite set
 is a special case of games with $\bH=\bL$,
 however, the latter condition allows also
 other types of games
 where the Learner can recover the losses of actions not chosen,
 and so which could be ``easier'' than classical bandits due to the knowledge of $\bL$.
Nevertheless, it is easy to see that these games are \emph{at most} as hard as classical bandit games.}
\begin{center}
\begin{minipage}[c]{5cm}
\begin{align*}
\bL & = \begin{pmatrix} 0 & 0 \\ -1 & 1 \end{pmatrix}, &
\bH & = \begin{pmatrix} 0 & 0 \\ -1 & 1 \end{pmatrix},
\end{align*}
\end{minipage}
\begin{minipage}[c]{9cm}
\includegraphics{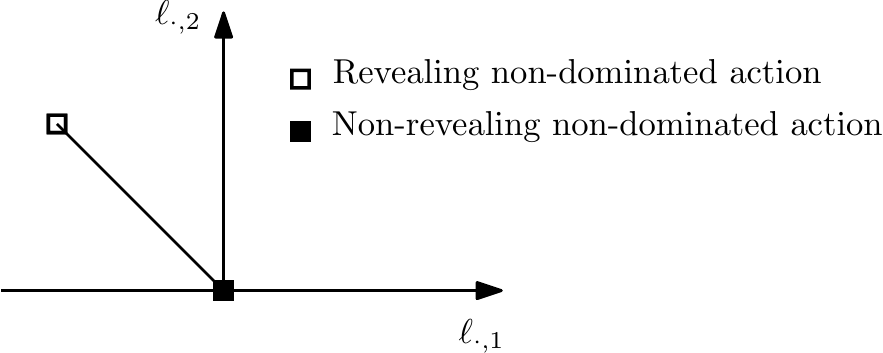}
\end{minipage}
\end{center}
Because the loss of the first action is $0$ regardless of the outcome,
 and the loss varies only for the second action, we call this game a \emph{one-armed bandit} game.
Both actions are non-dominated and the second one is revealing,
 therefore it is an easy game and according to Theorem~\ref{theorem:characterization}
 its minimax regret is $\widetilde \Theta(\sqrt{T})$.
(For this specific game, it can be shown that it is in fact $\Theta(\sqrt{T})$.)
\end{example}
\begin{example}[Apple Tasting]\label{ex:apple}
Consider an orchard that wants to hand out its crop of apples for sale.
However, some of the apples might be rotten.
The orchard can do a sequential test.
Each apple can be either tasted (which reveals whether the apple is healthy
 or rotten) or the apple can be given out for sale.
If a rotten apple is given
 out for sale, the orchard suffers a unit loss.
On the other hand, if a
 healthy apple is tasted, it cannot be sold and, again, the orchard suffers a unit loss.
This can be
 formalized by the following partial-monitoring game~\cite{Helmbold-Littlestone-Long-2000}:
\begin{center}
\begin{minipage}[c]{5cm}
\begin{align*}
\bL & = \begin{pmatrix} 1 & 0 \\ 0 & 1 \end{pmatrix}, &
\bH & = \begin{pmatrix} a & a \\ b & c \end{pmatrix},
\end{align*}
\end{minipage}
\begin{minipage}[c]{9cm}
\includegraphics{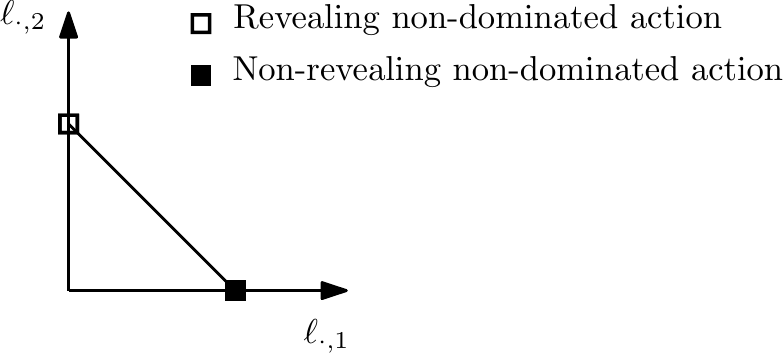}
\end{minipage}
\end{center}
The first action corresponds to giving out the apple for sale,
 the second corresponds to tasting the apple; the first outcome corresponds
 to a rotten apple, the second outcome corresponds to a healthy apple.
Both actions are non-dominated and the second one is revealing, therefore
 it is an easy game and according to Theorem~\ref{theorem:characterization} the minimax
 regret is $\widetilde \Theta(\sqrt{T})$. This is apparently a new result for this game.
Also notice that the picture is a just a translation of the picture for the one-armed bandit.
\end{example}
\begin{example}[Label Efficient Prediction]
Consider a situation when we would like to sequentially classify emails as spam or as legitimate.
For each email we have to output a prediction, and additionally
 we can request, as feedback, the correct label from the user.
If we classify an email incorrectly or we request its label, we suffer a unit loss.
(If the email is classified correctly and we do not request the feedback, no loss is suffered.)
This can be formalized by the following partial-monitoring game~\citep{Cesa-Bianchi-Lugosi-Stoltz-2006}:
\begin{center}
\begin{minipage}[c]{5cm}
\begin{align*}
\bL & = \begin{pmatrix} 1 & 1 \\ 0 & 1 \\ 1 & 0 \end{pmatrix}, &
\bH & = \begin{pmatrix} a & b \\ c & c \\ d & d \end{pmatrix},
\end{align*}
\end{minipage}
\begin{minipage}[c]{10cm}
\includegraphics{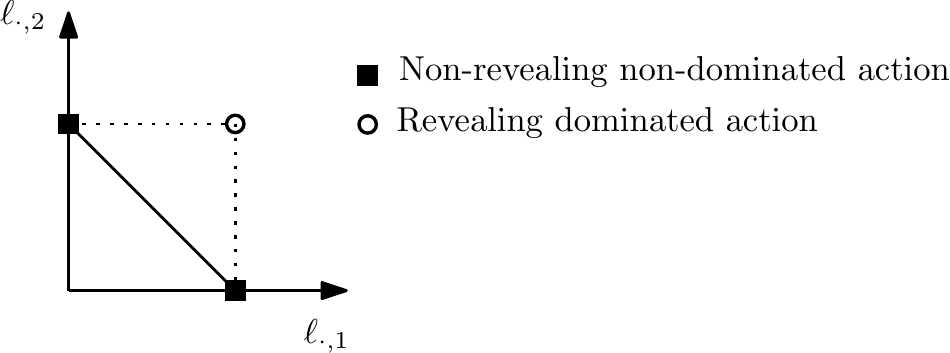}
\end{minipage}
\end{center}
where the first action corresponds to a label request, and the second and the
 third action correspond to a prediction (spam and legitimate, respectively)
 without a request.
The outcomes correspond to spam and legitimate emails.

We see that the chain of non-dominated actions contains two neighboring
 \nonrev\ actions and there is a dominated revealing action.
Therefore, it is a hard game and, by
 Theorem~\ref{theorem:characterization}, the minimax regret is $\Theta(T^{2/3})$.
This specific example was the only game known so far with minimax regret
 at least $\Omega(T^{2/3})$~\citep[Theorem~5.1]{Cesa-Bianchi-Lugosi-Stoltz-2006}.
\end{example}
\begin{example}[A Hopeless Game]
The following game is an example where the feedback does not reveal any information about the outcome:
\begin{center}
\begin{minipage}[c]{5cm}
\begin{align*}
\bL & = \begin{pmatrix} 1 & 0 \\ 0 & 1 \end{pmatrix}, &
\bH & = \begin{pmatrix} a & a \\ b & b \end{pmatrix},
\end{align*}
\end{minipage}
\begin{minipage}[c]{9cm}
\includegraphics{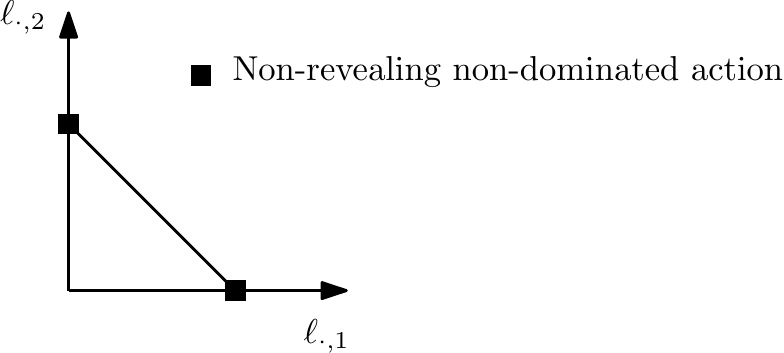}
\end{minipage}
\end{center}
Because both actions are \nonrev\ and non-dominated,
 it is a hopeless game and thus its minimax regret is $\Theta(T)$.
\end{example}
\begin{example}[A Trivial Game]
In the following game, the best action, regardless of the outcome sequence, is action $2$.
A learner that chooses this action in every round is guaranteed to have zero regret.
\begin{center}
\begin{minipage}[c]{5cm}
\begin{align*}
\bL & = \begin{pmatrix} 2 & 1 \\ 1 & 0 \\ 1 & 1 \end{pmatrix}, &
\bH & = \begin{pmatrix} a & b \\ c & d \\ e & f \end{pmatrix}
\end{align*}
\end{minipage}
\begin{minipage}[c]{11cm}
\includegraphics{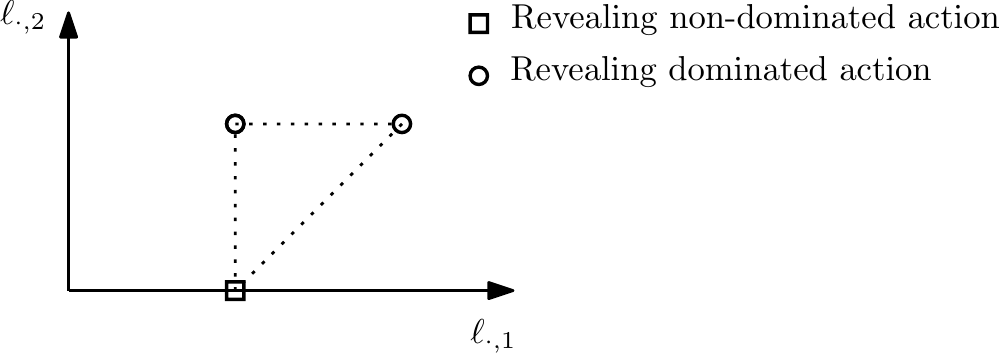}
\end{minipage}
\end{center}
Because this game has only one non-dominated action (action $2$),
 it is a trivial game and thus its minimax regret is $0$.
\end{example}
\begin{example}[A Degenerate Game]
The next game does not satisfy the non-de\-gen\-er\-a\-cy condition and therefore
 Theorem~\ref{theorem:characterization} does not apply.
\begin{center}
\begin{minipage}[c]{5cm}
\begin{align*}
\bL & = \begin{pmatrix} 2 & 0 \\ 1 & 1 \\ 0 & 2 \end{pmatrix}, &
\bH & = \begin{pmatrix} a & a \\ b & c \\ d & d \end{pmatrix}
\end{align*}
\end{minipage}
\begin{minipage}[c]{11cm}
\includegraphics{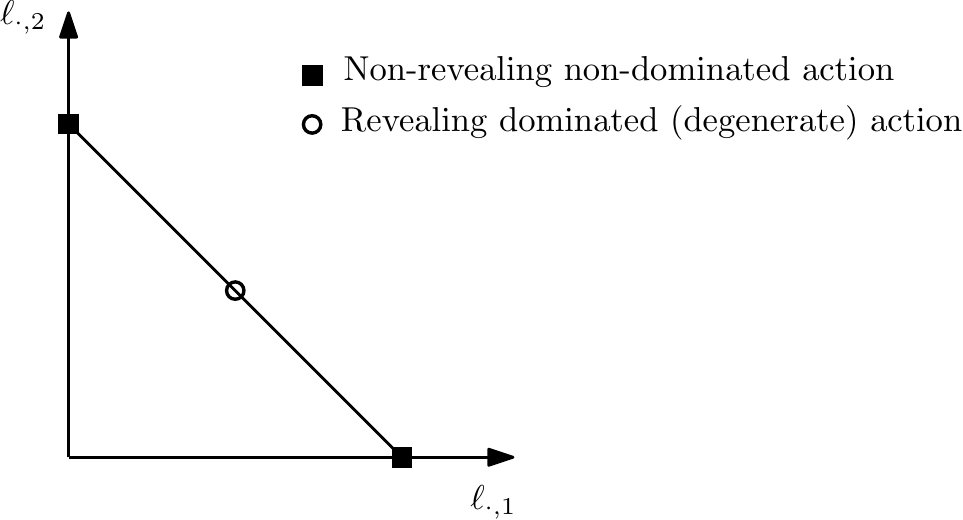}
\end{minipage}
\end{center}
Its minimax regret is between $\Omega(\sqrt{T})$ and $O(T^{2/3})$.
It remains an open problem to close this gap and determine the exact rate of growth.
\end{example}

\section{Upper bound for easy games}
\label{section:upper-bound}
In this section we present our algorithm for games satisfying the
 separation condition and the non-de\-gen\-er\-a\-cy condition, and prove that
 it achieves $\widetilde O(\sqrt{T})$ regret with high probability.
We call the algorithm \AppleTree\ since it builds
 a binary tree, leaves of which are apple tasting games.

\subsection{The algorithm}
\label{section:algorithms}
In the first step of the algorithm we can purify the game by first removing the
 dominated actions and then the duplicates as mentioned beforehand.

The idea of the algorithm is to recursively split the game until we arrive at
 games with two actions only.
Now, if one has only two actions in a
 partial-information game, the game must be either a full-information game (if
 both actions are revealing) or an instance of a one-armed bandit (with one revealing and one \nonrev\ action).

To see why this latter case corresponds to one-armed bandits,
 assume without loss of generality that the first action is the revealing action.
Now, it is easy to see that
 the regret of a sequence of actions in a game does not change if the loss matrix is
 changed by subtracting the same number from a column.%
\footnote{As a result, for any algorithm,
 if $\Regret_T$ is its regret at time $T$ when measured in the game
 with the modified loss matrix, the algorithm's ``true'' regret will also be
 $\Regret_T$ (\ie the algorithm's regret when measured in the original, unmodified game).
\citet{Piccolboni-Schindelhauer-2001} exploit this idea, too.}
By subtracting $\ell_{2,1}$ from the first and $\ell_{2,2}$ from the second column
we thus get the equivalent game where the second row of the loss matrix is zero,
arriving at a one-armed bandit game (see Example~\ref{ex:one-armed}).
Since a one-armed
bandit is a special form of a two-armed bandit, one can use Exp3.P due to
\citet{Auer-Cesa-Bianchi-Freund-Schapire-2002} to achieve the $O(\sqrt{T})$ regret.

\begin{figure}[t]
\centering
  \includegraphics[scale=1]{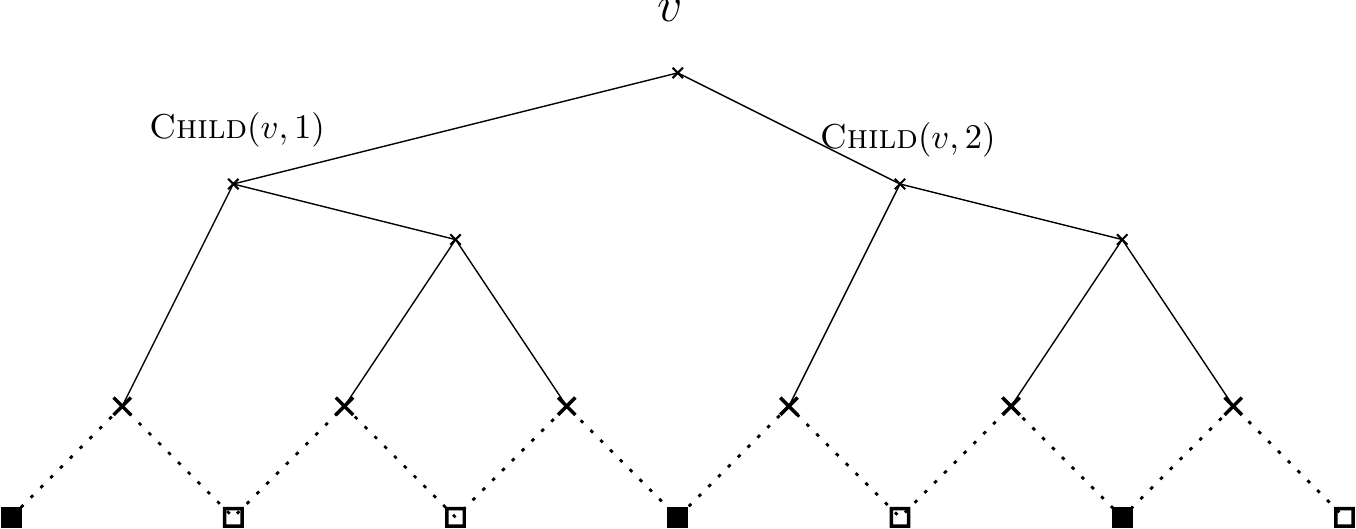}
  \caption{The binary tree built by the algorithm. The leaf nodes represent neighboring action pairs.}
\label{fig:tree}
\end{figure}

Now, if there are more than two actions in the game, then the game is split,
putting the first half of the actions into the first and the second half into
the second subgame, with a \emph{single common shared action}. Recall that, in the chain of non-dominated actions, the
actions are ordered according to their losses corresponding to the {\em first} outcome.
This is continued until the split results in games with two actions only.
The recursive splitting of the game results in a binary tree (see Figure~\ref{fig:tree}).
The idea of the strategy played at an internal node of the tree is as follows:
An outcome sequence of length $T$ determines the frequency $\rho_T$ of outcome
$2$. If this frequency is small, the optimal action is one of the actions of
$G_1$, the first subgame (simply because then the frequency of outcome $1$ is
high and $G_1$ contains the actions with the smallest loss for the first
outcome). Conversely, if this frequency is large, the optimal action is one of
the actions of $G_2$. In some intermediate range, the optimal action is the
action shared between the subgames. Let the boundaries of this range be
$\rho_1^*<\rho_2^*$ ($\rho_1^*$ is thus the solution to $(1-\rho)\ell_{s-1,1} +
\rho\ell_{s-1,2} = (1-\rho)\ell_{s,1} + \rho \ell_{s,2}$ and $\rho_2^*$ is the
solution to $(1-\rho)\ell_{s+1,1} + \rho\ell_{s+1,2} = (1-\rho)\ell_{s,1} +
\rho \ell_{s,2}$, where $s = \lceil K/2\rceil$ is the index of the action
shared between the two subgames.)

If we knew $\rho_T$, a good solution would be to play a strategy where the
actions are restricted to that of either game $G_1$ or $G_2$, depending on
whether $\rho_T\le \rho_1^*$ or $\rho_T\ge \rho_2^*$. (When $\rho_1^* \le
\rho_T \le \rho_2^*$ then it does not matter which action-set we restrict the
play to, since the optimal action in this case is included in both sets.) There
are two difficulties. First, since the outcome sequence is not known in
advance, the best we can hope for is to know the running frequencies
$
\rho_t = \frac{1}{t} \,\sum_{s=1}^t \one{J_s=2}
$.
However, since the game is a partial-information game, the outcomes are not
revealed in all time steps, hence, even $\rho_t$ is inaccessible.
Nevertheless, for now let us assume that $\rho_t$ was available. Then one
idea would be to play a strategy restricted to the actions of either game $G_1$
or $G_2$ as long as $\rho_t$ stays below $\rho_1^*$ or above $\rho_2^*$.
Further, when $\rho_t$ becomes larger than $\rho_2^*$ while previously the
strategy played the action of $G_1$ then we have to switch to the game $G_2$.
In this case, we start a fresh copy (a \emph{reset}) of a strategy playing in $G_2$. The same
happens when a switch from $G_2$ to game $G_1$ is necessary. These resets are
necessary because at the leaves we play according to strategies that use
weights that depend on the cumulated losses of the actions
\emph{exponentially}. To see an example when without resets the algorithm
fails to achieve a small regret consider the case when there are $3$ actions,
the middle one being revealing. Assume that during the first $T/2$ time steps
the frequency of outcome $2$ oscillates between the two boundaries so that the
algorithm switches constantly back and forth between the games $G_1$ and $G_2$.
Assume further that in the second half of the game, the outcome is always $2$.
This way the optimal action will be $3$. Nevertheless, up to time step $T/2$,
the player of $G_2$ will only see outcome $1$ and thus will think that action
$2$ is the optimal action. In the second half of the game, he will not have
enough time to recover and will play action $2$ for too long. Resetting the algorithms of the
subgames avoids this behavior.

If the number of switches was large, the repeated resetting of the strategies
could be equally problematic. Luckily this cannot happen, hence the resetting
does minimal harm. We will in fact show that this generalizes to the case even
when $\rho_t$ is estimated based on partial feedback (see Lemma~\ref{lem:resets}).

Let us now turn to how $\rho_t$ is estimated.
As mentioned in Section~\ref{section:characterization},
 mapping a row of $\bH$ bijectively leads to an equivalent game,
 thus for $M=2$ we can assume without loss of generality that
 in any round, the algorithm receives (possibly random) feedback $H_t \in \{1,2,*\}$:
 if a revealing action is played in the round, $H_t = J_t \in \{1,2\}$,
 otherwise $H_t = *$.
Let $\H_{1:t-1} = (I_1,H_1,\dots,I_{t-1},H_{t-1}) \in (\LAs\times\Sigma)^{t-1}$,
 the (random) history of actions and observations up to time step $t-1$.
If the algorithm choosing the actions decides with probability $p_t\in (0,1]$ to play a revealing action
($p_t$ can depend on $\H_{1:t-1}$) then $\one{H_t=2}/p_t$ is a simple
unbiased estimate of $\one{J_t=2}$ (in fact, $\Exp \left[ \one{H_t=2}/p_t | \H_{1:t-1} \right] =
\one{J_t=2}$). As long as $p_t$ does not drop to a too low value,
$
\hr_t = \frac{1}{t} \, \sum_{s=1}^t \frac{\one{H_s=2}}{p_s}
$
will be a relatively reliable estimate of $\rho_t$ (see Lemma~\ref{lem:highprob}).
However reliable this estimate is, it can still differ from $\rho_t$. For this reason, we push the boundaries determining game switches towards each other:
\begin{align}
\label{def:rho}
	\rho'_1=\frac{2\rho^*_1+\rho^*_2}{3}, \quad
	\rho'_2=\frac{\rho^*_1+2\rho^*_2}{3}.
\end{align}

We call the resulting algorithm \AppleTree, because the elementary
partial-information $2$-action games in the bottom essentially correspond to
instances of the apple tasting problem (see Example~\ref{ex:apple}).
The algorithm's main entry point is shown on \Algfig~\ref{alg:main}. Its
inputs are the game $G=(\bL,\bH)$, the time horizon and a confidence parameter
$0<\delta<1$. The algorithm first eliminates the dominated and duplicate
actions. This is followed by building a tree, which is used to store variables
necessary to play in the subgames (\Algfig~\ref{alg:buildTree}):
If the number
of actions is $2$, the procedure initializes various parameters that are used
either by a bandit algorithm (based on Exp3.P
\citep{Auer-Cesa-Bianchi-Freund-Schapire-2002}), or by the exponentially
weighted average algorithm (EWA)~\citep{Cesa-Bianchi-Freund-Haussler-Helmbold-Schapire-Warmuth-1997}. In the other case, it calls
itself recursively on the split subgames and with an appropriately decreased
confidence parameter.

\begin{figure}[tb]
\twocolumns{0.42}{0.54}{t}{t}{
	\begin{algorithmic}[1]
	\Statex \mbox{} \hspace*{-2em} {\bf function} \Call{Main}{$G,T,\delta$}
	\Statex \mbox{} \hspace*{-2em} \textbf{Input:} $G=(\bL,\bH)$ is a game, $T$ is a horizon, $0<\delta<1$ is a confidence parameter
	\State $G \gets \Call{Purify}{G}$
	\State $\Call{BuildTree}{ \mathrm{\bf root}, G,\delta }$
	\For{$t\gets 1$ \algorithmicto\ $T$}
		\State $\Call{Play}{\mathrm{\bf root}}$
	\EndFor
	\end{algorithmic}
	\caption{The main entry point of the {\sc AppleTree} algorithm}
	\label{alg:main}
	\vspace*{0.2in}
	\begin{algorithmic}[1]
	\Statex \mbox{} \hspace*{-2em} {\bf function} \Call{InitEta}{$G,T$}
	\Statex \mbox{} \hspace*{-2em} \textbf{Input:} $G$ is a game, $T$ is a horizon
	\If{ $\Call{IsRevealing}{G,2}$}
		\State $\eta(v) \gets \sqrt{8 \ln2\, /T}$
	\Else
		\State $\eta(v) \gets \gamma(v)/4$
	\EndIf
	\end{algorithmic}
	\caption{The initialization routine {\sc InitEta}.}
}{
	\vspace*{-0.10in}
	\begin{algorithmic}[1]
	\Statex \mbox{} \hspace*{-2em} {\bf function} \Call{BuildTree}{$v,G,\delta$}
	\Statex \mbox{} \hspace*{-2em} \textbf{Input:} $G=(\bL,\bH)$ is a game, $v$ is a tree node
	\If{$\Call{NumOfActions}{G}=2$}
		\If{ {\bf not} $\Call{IsRevealing}{G,1}$}
			\State $G\gets \Call{SwapActions}{G}$
		\EndIf
		\State $w_i(v) \gets 1/2$, $i=1,2$
			\State $\beta(v) \gets \sqrt{\ln(2/\delta)/(2T)}$
		\State $\gamma(v) \gets 8\beta(v)/(3+\beta(v))$
		\State \Call{InitEta}{$G,T$}
	\Else
		\State $(G_1,G_2) \gets \Call{SplitGame}{G}$
		\State \Call{BuildTree}{{\sc Child}($v,1$), $G_1,\delta/(4T)$ }
		\State \Call{BuildTree}{{\sc Child}($v,2$), $G_2,\delta/(4T)$ }
		\State $g(v) \gets 1$, $\hr(v) \gets 0$, $t(v)\gets 1$
		\State $(\rho_1'(v),\rho_2'(v))\gets \Call{Boundaries}{G}$
	\EndIf
	\State $G(v) \gets G$
	\end{algorithmic}
	\vspace*{.42in}
	\caption{\label{alg:buildTree} The tree building procedure}

}
\end{figure}

The main worker routine is called {\sc Play}. This is again a recursive
function (see \Algfig~\ref{alg:play}). The special case when the number of
actions is two is handled in routine {\sc PlayAtLeaf}, which will be discussed
later. When the number of actions is larger, the algorithm recurses to play in
the subgame that was remembered as the game to be preferred from the last round
and then updates its estimate of the frequency of outcome $2$ based on the
information received. When this estimate changes so that a switch of the
current preferred game is necessary, the algorithm resets the algorithms in the
subtree corresponding to the game switched to, and changes the variable storing
the index of the preferred game. The {\sc Reset} function used for this
purpose, shown on Figure~\ref{alg:reset}, is also recursive.

\begin{figure}[tb]
\twocolumns{0.55}{0.42}{t}{t}{
	\begin{algorithmic}[1]
	\Statex \mbox{} \hspace*{-2em} {\bf function} \Call{Play}{$v$}
	\Statex \mbox{} \hspace*{-2em} \textbf{Input:} $v$ is a tree node
		\If{$ \Call{NumOfActions}{G(v)}=2$}
		\State $(p,h)\gets 	\Call{PlayAtLeaf}{v}$
	\Else
		\State $(p,h)\gets $	\Call{Play}{{\sc Child}($v,g(v)$)}
		\State $\hr(v) \gets (1-\frac{1}{t(v)}) \hr(v) + \frac{1}{t(v)} \, \frac{\one{h=2}}{p}$
		\If{$g(v)=2$ {\bf and} $\hr(v)<\rho_1'(v)$}
			\State \Call{Reset}{{\sc Child}($ v, 1 $)}; $g(v) \gets 1$
		\ElsIf{$g(v)=1$ {\bf and} $\hr(v)>\rho_2'(v)$}
			\State \Call{Reset}{{\sc Child}($ v, 2 $)}; $g(v) \gets 2$
		\EndIf
		\State $t(v)\gets t(v)+1$
	\EndIf
	\State  \Return $(p,h)$
	\end{algorithmic}
\caption{\label{alg:play} The recursive function {\sc Play}}
}
{
	\begin{algorithmic}[1]
	\Statex \mbox{} \hspace*{-2em} {\bf function} \Call{Reset}{$v$}
	\Statex \mbox{} \hspace*{-2em} \textbf{Input:} $v$ is a tree node
	\If{$\Call{NumOfActions}{G(v)}=2$}
			\State $w_i(v) \gets 1/2$, $i\gets 1,2$
	\Else
		\State $g(v)\gets 1 $, $\hr(v) \gets 0 $, $t(v) \gets 1$
		\State \Call{Reset}{{\sc Child}($v,1$)}
	\EndIf
	\end{algorithmic}
	\vspace*{.96in}
\caption{\label{alg:reset} Function {\sc Reset}}
}
\end{figure}

At the leaves, when there are only two actions, either EWA or Exp3.P is used.
These algorithms are used with their standard optimized parameters (see
Corollary~4.2 for the tuning of EWA, and Theorem~6.10 for the tuning of Exp3.P,
both from the book of \citet{Cesa-Bianchi-Lugosi-2006}). For completeness,
their pseudocodes are shown in \Algfigs~\ref{alg:playatleaf}--\ref{alg:EWA}.
Note that with Exp3.P (lines~\ref{line:exp3.p.start}--\ref{line:exp3.p.end}) we
use the loss matrix transformation described earlier, hence the loss matrix has
zero entries for the second (\nonrev) action, while the entry for action
$1$ and outcome $j$ is $\ell_{1,j}(v)-\ell_{2,j}(v)$. Here $\ell_{i,j}(v)$
stands for the loss of action $i$ and outcome $j$ in the game $G(v)$ that is
stored at node $v$.

\renewcommand{\And}{{\bf and}}
\begin{figure}[tb]
\twocolumns{0.53}{0.44}{b}{b}{
	\begin{algorithmic}[1]
	\Statex \mbox{} \hspace*{-2em} {\bf function} \Call{PlayAtLeaf}{$v$}
	\Statex \mbox{} \hspace*{-2em} \textbf{Input:} $v$ is a tree node
	\If{$\Call{RevealingActionNumber}{G(v)}=2$} \Comment{Full-information case}
			\State $(p,h) \gets \Call{Ewa}{v}$ 
	\Else \Comment{Partial-information case}
		\State $p \gets (1-\gamma(v))\frac{w_1(v)}{w_1(v)+w_2(v)} + \gamma(v)/2$
		\State $U \sim {\cal U}_{[0,1)}$ \Comment{$U$ is uniform in $[0,1)$}
		\If{$U<p$} \Comment{Play revealing action}\label{line:exp3.p.start}
			\State $h\gets \mathrm{CHOOSE}( 1 )$ \Comment{$h\in \{1,2\}$}
			\State $L_1 \gets( \ell_{1,h}(v) - \ell_{2,h}(v) + \beta(v))/p$
			\State $L_2 \gets \beta(v)/(1-p)$
			\State $w_1(v) \gets w_1(v) \exp( -\eta(v) L_1 )$
			\State $w_2(v) \gets w_2(v) \exp( -\eta(v) L_2 )$
		\Else
			\State $h\gets \mathrm{CHOOSE}( 2 )$ \Comment{here $h=*$}
		\EndIf \label{line:exp3.p.end}
	\EndIf
	\State  \Return $(p,h)$
	\end{algorithmic}
	\caption{Function {\sc PlayAtLeaf}}
	\label{alg:playatleaf}
}{
	\begin{algorithmic}[1]
	\Statex \mbox{} \hspace*{-2em} {\bf function} \Call{Ewa}{$v$}
	\Statex \mbox{} \hspace*{-2em} \textbf{Input:} $v$ is a tree node
	\State $p \gets \frac{w_1(v)}{w_1(v)+w_2(v)}$
	\State $U \sim {\cal U}_{[0,1)}$ \Comment{$U$ is uniform in $[0,1)$}
	\If{$U<p$}
		\State $I\gets 1$
	\Else
		\State $I\gets 2$
	\EndIf
	\State $h\gets \mathrm{CHOOSE}( I )$ \Comment{$h\in \{1,2\}$}
	\State $w_1(v) \gets w_1(v) \exp( -\eta(v) \ell_{1,h}(v) )$
	\State $w_2(v) \gets w_2(v) \exp( -\eta(v) \ell_{2,h}(v) )$
	\State  \Return $(p,h)$
	\end{algorithmic}
	\vspace*{0.9in}
	\caption{Function {\sc Ewa}}
	\label{alg:EWA}
}
\end{figure}

\subsection{Proof of the upper bound}
\begin{theorem}
\label{theorem:sqrt-upper}
Assume $G=(\bL,\bH)$ satisfies the separation condition and the
non-de\-gen\-er\-a\-cy condition and $\ell_{i,j}\leq1$. Denote by
$\hRegret_T$ the regret of Algorithm \AppleTree{} up to time step $T$.
There exist constants $c$,$p$ such that for any $0<\delta<1$ and $T\in\N$,
 for any outcome sequence $J_1,\dots,J_T$,
 the algorithm with input $G,T,\delta$ achieves
$
\Pr\left[\hRegret_T\leq c\sqrt{T}\ln^p(2T/\delta)\right]\geq1-\delta \;.
$
\end{theorem}

Throughout the proof we will analyze the algorithm's behavior at the root node.
We will use time indices as follows. Let us define the filtration
$\{\F_t=\sigma(I_1,\dots,I_t)\}_t$, where $I_t$ is the action the
algorithm plays at time step $t$. To any variable $x(v)$ used by the algorithm,
we denote by $x_t(v)$ the value of $x(v)$ that is measurable with respect to
$\F_t$, but not measurable with respect to $\F_{t-1}$. From
now on we abbreviate $x_t(\mathrm{root})$ by $x_t$. We start with two lemmas.
The first lemma shows that the number of switches the algorithm makes is small.
\begin{lemma}
\label{lem:resets}
Let $S$ be the number of times \AppleTree{} calls \textsc{Reset} at the
root node. Then there exists a universal constant $c^*$ such that $S\leq
\frac{c^*\ln T}{\Delta}$, where $\Delta=\rho'_2-\rho'_1$ with $\rho'_1$ and
$\rho'_2$ given by~\eqref{def:rho}.
\end{lemma}

Note that here we use the non-de\-gen\-er\-a\-cy condition to ensure that $\Delta > 0$.

\begin{proof}
Let $s$ be the number of times the algorithm switches from $G_2$ to $G_1$. Let
$t_1<\dots<t_s$ be the time steps when $\hr_t$ becomes smaller than
$\rho'_1$. Similarly, let $t'_1<\dots<t'_{s+\xi},\ (\xi\in\{0,1\})$ be the
time steps when $\hr_t$ becomes greater than $\rho'_2$. Note that for
all $1\leq j<s,\ t'_j<t_j<t'_{j+1}$. Finally, for every $1\leq j<s$, we define $t''_j=\min\{t~|~t'_j\leq t \leq t_j, (\forall t\leq\tau\leq t_j: \hr_\tau\leq1)\}$. In other words, $t''_j$ is the time step when $\hr_t$ drops below $1$ and stays there until the next reset.

First we observe that if $t''_j\geq2/\Delta$ then $\hr_{t''_j}\geq(\rho'_1+\rho'_2)/2$. Indeed, if $t''_j=t'_j$ then $\hr_{t''_j}\geq\rho'_2$, on the other hand, if $t''_j\neq t'_j$ then $\hr_{t''_j-1}>1$ and, from the update rule we have
\begin{align*}
  \hr_{t''_j}=\left(1-\frac{1}{t''_j}\right)\hr_{t''_j-1} +\frac{1}{t''_j}\cdot\frac{\one{J_{t''_j}=2}}{p_{t''_j}} \geq1-\frac\Delta2\geq\frac{\rho'_1+\rho'_2}{2} \;.
\end{align*}

The number of times the algorithm resets
is at most $2s+1$. Let $j^*$ be the first index such that $t''_{j^*}\geq2/\Delta$. For any $j^*\leq j\leq s$, $\hr_{t''_j}\geq(\rho'_1+\rho'_2)/2$ and
$\hr_{t_j}\leq\rho'_1$. According to the update rule we have for any $t''_j<t\leq t_j$
that
\begin{align*}
 \hr_t =
 \left(1-\frac{1}{t}\right)\hr_{t-1}+\frac{1}{t}\cdot\frac{\one{J_t=2}}{p_{t}}
 \geq \hr_{t-1}-\frac{1}{t}\hr_{t-1}\geq \hr_{t-1}-\frac{1}{t}
\end{align*}
and hence
$
 \hr_{t-1}-\hr_t\leq\frac{1}{t} \;.
$
Summing this inequality for all $t''_j+1\leq t\leq t_j$ such that $j\geq j^*$ we get
\begin{align*}
 \frac{\Delta}{2}=\frac{\rho'_1+\rho'_2}{2}-\rho'_1&\leq\hr_{t''_j}-\hr_{t_j}\\
 &\leq\sum_{t=t''_j+1}^{t_j}\frac{1}{t}= O\left(\ln\frac{t_j}{t''_j}\right) \;.
\end{align*}
Thus, there exists $c>0$ such that for all $j^*\leq j\leq s$
\begin{align}
\label{eq:to-add} \frac{1}{c}\Delta\leq\ln\frac{t_j}{t''_j}\leq\ln\frac{t_j}{t_{j-1}} \;.
\end{align}
Adding~\eqref{eq:to-add} for $j^*< j\leq s$ we get
$
 (s-j^*)\frac{1}{c}\Delta\leq\ln\frac{t_s}{2/\Delta}\leq\ln T \;.
$
We conclude the proof with observing that $j^*\leq2/\Delta$.
\end{proof}

The next lemma shows that the estimate of the relative frequency of outcome 2
is not far away from its true value.
\begin{lemma}
\label{lem:highprob} 
For any $0<\delta<1$, with
probability at least $1-\delta$, for all $t\geq 8\sqrt{T}\ln (2T/\delta)/(3\Delta^2)$,
$|\hr_t-\rho_t|\leq\Delta$.
\end{lemma}

The proof of the lemma employs Bernstein's inequality for martingales.
\begin{bernstein}~\cite[Lemma A.8]{Cesa-Bianchi-Lugosi-2006}
Let $X_1, X_2, \dots, X_n$ be a bounded martingale difference
sequence with respect to a filtration $\{\F\}_{i=0}^n$ and with $|X_i| \le K$.
Let
$$
S_i = \sum_{j=1}^i X_j
$$
be the associated martingale. Denote the sum of conditional variances by
$$
\Sigma_{n}^2 = \sum_{i=1}^n \Exp[X_i^2 ~|~ \F_{i-1} ] \;.
$$
Then, for all constants $\epsilon,v > 0$,
$$
\Pr \left[ \max_{i\in\underline{n}} S_i > \epsilon \ \text{and} \ \Sigma_n^2 \le v \right] \le \exp\left( - \frac{\epsilon^2}{2(v + K\epsilon/3)} \right) \;.
$$
\end{bernstein}

\begin{proof}[Proof of Lemma~\ref{lem:highprob}]
For $1\leq t\leq T$, let $p_t$ be
 the conditional probability of playing a revealing action at time step $t$,
 given the history $\H_{1:t-1}$.
Recall that, due to the construction of the algorithm, $p_t\geq1/\sqrt{T}$.

If we write $\hr_t$ in its explicit form
$\hr_t=\frac{1}{t}\sum_{s=1}^t\frac{\one{H_s=2}}{p_s}$
we can observe that $\Exp[\hr_t|\H_{1:t-1}]=\rho_t$, that is, $\hr_t$
is an unbiased estimate of the relative frequency. Let us define random
variables $X_s:=\frac{\one{H_s=2}}{p_s}-\one{J_s=2}$.
Since $p_s$ is determined by the history, $\{X_s\}_s$ is a martingale
difference sequence. Also, from $p_s\geq1/\sqrt{T}$ we know that $\Var(X_s|\H_{1:t-1})\leq\sqrt{T}$.
Hence, we can use Bernstein's inequality for martingales
with $\epsilon=\Delta t$, $\nu=t\sqrt{T}$, $K=\sqrt{T}$:
\begin{align*}
\Pr \left[|\hr_t-\rho_t|>\Delta \right]
& = \Pr \left[ \left|\sum_{s=1}^t X_s \right| >t \Delta \right]\\
& \le 2 \exp\left(-\frac{\Delta^2t^2/2}{t\sqrt{T}+\Delta t\sqrt{T}/3}\right)\\
& \le 2 \exp\left(-\frac{3\Delta^2t}{8\sqrt{T}}\right) \;.
\end{align*}
We have that if $t\geq 8\sqrt{T}\ln (2T/\delta)/(3\Delta^2)$ then
\begin{align*}
\Pr \left[|\hr_t-\rho_t|>\Delta\right]\leq \delta/T \;.
\end{align*}
We get the bound for all $t\in [8\sqrt{T}\ln (2T/\delta)/(3\Delta^2),T]$ using the union bound.

\end{proof}

\begin{proof}[Proof of Theorem~\ref{theorem:sqrt-upper}]
To prove that the algorithm achieves the desired regret bound we use induction
on the depth of the tree, $d$. If $d=1$, \AppleTree{} plays either EWA or
Exp3.P. EWA is known to satisfy Theorem~\ref{theorem:sqrt-upper}, and, as we
discussed earlier, Exp3.P achieves $O(\sqrt{T}\ln T/\delta)$ regret as well. As
the induction hypothesis we assume that Theorem~\ref{theorem:sqrt-upper} is
true for any $T$ and any game such that the tree built by the algorithm has
depth $d'<d$.

Let $Q_1=\{1,\dots,\lceil K/2\rceil\}$, $Q_2=\{\lceil K/2\rceil,\dots,K\}$ be
the sets of actions associated with the subgames in the root.
(Recall that the actions are ordered with respect to $\ell_{\cdot,1}$.)
Furthermore,
let us define the following values:
Let $T_0^0=1$, let $T_i^0$ be the first
time step $t$ after $T_{i-1}^0$ such that $g_t\neq g_{t-1}$. In other words,
$T_i^0$ are the time steps when the algorithm switches between the subgames.
Finally, let $T_i=\min(T_i^0,T+1)$. From Lemma~\ref{lem:resets} we know that
$T_{\Smax+1}=T+1$, where $\Smax=\frac{c^*\ln T}{\Delta}$. It is easy to see
that $T_i$ are stopping times for any $i\geq1$.

Without loss of generality, from now on we will assume that the optimal action $i^*\in Q_1$.
If $i^*=\lceil K/2\rceil$ then, since it is contained in both subgames,
the bound trivially follows from the induction hypothesis and Lemma~\ref{lem:resets}.
In the rest of the proof we assume $i^*<K/2$.

Let
$S = \max\{i \geq 1 ~|~ T_i^0\leq T\}$ be the number of switches,
 $c=\tfrac{8}{3\Delta^2}$, and $\B$ be
the event that for all $t\geq c\sqrt{T}\ln (4T/\delta)$,
$|\hr_t-\rho_t|\leq\Delta$. We know from Lemma~\ref{lem:highprob} that
$\Pr[\B]\geq 1-\delta/2$. On $\B$ we have that
$|\hr_T-\rho_T|\leq\Delta$, and thus, using that $i^*<K/2$,
$\rho_{T}\leq\rho_1^*$. This implies that in the last phase the algorithm plays
on $G_1$. It is also easy to see that before the last switch, at time step
$T_{S}-1$, $\hr$ is between $\rho_1^*$ and $\rho_2^*$, if $T_S$ is large
enough. Thus, up to time step $T_S-1$, the optimal action is $\lceil
K/2\rceil$, the one that is shared by the two subgames. This implies that
$
 \sum_{t=1}^{T_S-1}\ell_{i^*,J_t}-\ell_{\lceil K/2\rceil,J_t} \ge 0.
$
On the other hand, if $T_S\leq c\sqrt{T}\ln(4T/\delta)$ then
\begin{align*}
 \sum_{t=1}^{T_S-1}\ell_{i^*,J_t}-\ell_{\lceil K/2\rceil,J_t}\geq-c\sqrt{T}\ln(4T/\delta) \;.
\end{align*}
Thus, we have
\begin{align*}
\hRegret_T
& = \sum_{t=1}^T\ell_{I_t,J_t}-\ell_{i^*,J_t}\\
& = \sum_{t=1}^{T_S-1} \left(\ell_{I_t,J_t} - \ell_{i^*,J_t}\right) + \sum_{t=T_S}^T \left(\ell_{I_t,J_t}-\ell_{i^*,J_t}\right)\\
\begin{split}
& \leq \one{\B} \left(\sum_{t=1}^{T_S-1} \left( \ell_{I_t,J_t} - \ell_{\lceil K/2\rceil,J_t} \right)
 + \sum_{t=T_S}^T \left( \ell_{I_t,J_t}-\ell_{i^*,J_t} \right) \right)\\
& \quad + \underbrace{c\sqrt{T}\ln (4T/\delta)+\left(\one{\B^c}\right)T}_D
\end{split}\\
&\leq D + \one{\B}\sum_{r=1}^\Smax \max_{i\in Q_{\pi(r)}}\sum_{t=T_{r-1}}^{T_r-1} \left(\ell_{I_t,J_t}-\ell_{i,J_t}\right)\\
&=D+ \one{\B}\sum_{r=1}^\Smax \max_{i\in Q_{\pi(r)}}\sum_{m=1}^T\one{T_r-T_{r-1}=m} \sum_{t=T_{r-1}}^{T_{r-1}+m-1}\left(\ell_{I_t,J_t}-\ell_{i,J_t}\right)\;,
\end{align*}
where $\pi(r)$ is 1 if $r$ is odd and 2 if $r$ is even. Note that for the last
line of the above inequality chain to be well defined, we need outcome
sequences of length at most $2T$. It does us no harm to assume that for all
$T<t\leq2T$, say, $J_t=1$.

Recall that the strategies that play in the subgames are reset after the
switches. Hence, the sum
$\hRegret^{(r)}_{m}=\sum_{t=T_{r-1}}^{T_{r-1}+m-1}\left(\ell_{I_t,J_t}-\ell_{i,J_t}\right)$
is the regret of the algorithm if it is used in the subgame $G_{\pi(r)}$ for
$m\le T$ steps. Then, exploiting that $T_r$ are stopping times, we can use the
induction hypothesis to bound $\hRegret^{(r)}_m$. In particular, let
$\C$ be the event that for all $m\leq T$ the sum is less than
$c\sqrt{T}\ln^p(2T^2/\delta)$. Since the root node calls its children with
confidence parameter $\delta/(2T)$, we have that
$\Pr[\C^c]\leq\delta/2$. In summary,
\begin{align*}
 \hRegret_T&\leq D+\one{\C^c}T+ \one{\B}\one{\C}\Smax c\sqrt{T}\ln^p2T^2/\delta\\
&\leq \one{\B^c\cup\C^c}T+ c\sqrt{T}\ln(4T/\delta)+\one{\B}\one{\C}\frac{c^*\ln T}{\Delta}c\sqrt{T}\ln^p2T^2/\delta.
\end{align*}
Thus, on $\B\cap\C$,
$
 \hRegret_T\leq\frac{2^pcc^*}{\Delta}\sqrt{T}\ln^{p+1}\left(2T/\delta\right),
$
which, together with $\Pr[\B^c\cup\C^c]\leq\delta$ concludes the proof.
\end{proof}

\noindent\textbf{Remark}\quad
The above theorem proves a high probability bound on the regret.
We can get a bound on the expected regret if we set $\delta$ to $1/\sqrt{T}$.
Also note that the bound given by the induction grows in the number of
non-dominated actions as $O(K^{\log_2 K})$.

\section{Lower Bound for Non-Trivial Games}
\label{section:lower-bound-easy-games}
In the following sections,
 $\|\cdot\|_1$ and $\|\cdot\|$ denote
 the $L_1$- and $L_2$-norm of a vector in a Euclidean space, respectively.

In this section,
 we show that non-trivial games have minimax regret at least $\Omega(\sqrt{T})$.
We state and prove this result for \emph{all} finite games,
 in contrast to earlier related lower bounds which apply to specific losses
(see
 Cesa-Bianchi and Lugosi~\cite[Theorems~3.7, 6.3, 6.4, 6.11]{Cesa-Bianchi-Lugosi-2006} for full-information, label efficient, and bandit games).
\begin{theorem}[Lower bound for non-trivial games]
\label{theorem:sqrt_lower_bound_nontriv_games}
If $G=(\bL,\bH)$ is a finite non-trivial ($K\ge 2$) partial-monitoring game 
 then there exists a constant $c>0$ such that for any $T\ge 1$
 the minimax expected regret $\Regret_T(G)\ge c\sqrt{T}$.
\end{theorem}

The proof presented below works for stochastic nature, as well.
There is a far simpler proof in the Appendix, however, that one applies only for adversarial nature.

Recall that $\Delta_M\subset\R^M$ is
 the $(M-1)$-dimensional probability simplex.

For the proof, we start with a geometrical lemma,
 which ensures the existence of a pair $i_1$,$i_2$ of non-dominated actions
 that are ``neighbors'' in the sense
 that for any small enough $\epsilon>0$,
 there exists a pair of ``$\epsilon$-close'' outcome distributions $p + \epsilon v$ and $p - \epsilon v$ such
 that $i_1$ is uniquely optimal under the first distribution,
 and $i_2$ is uniquely optimal under the second distribution
 overtaking each non-optimal action by at least $\Omega(\epsilon)$ in both cases.
\begin{lemma}[$\epsilon$-close distributions]
\label{lem:epsilon-close-distributions}
Let $G=(\bL,\bH)$ be any finite non-trivial game with $N$ non-duplicate actions and $M\ge 2$ outcomes.
Then there exist two non-dominated actions $i_1$,$i_2\in\LAs$,
 $p \in \Delta_M$, $v \in \R^M\setminus\{0\}$, and $c$,$\alpha>0$
 satisfying the following properties:
\begin{itemize}
\item[(a)] $\ell_{i_1}\ne\ell_{i_2}$.
\item[(b)] $\langle \ell_{i_1},p \rangle = \langle \ell_{i_2},p \rangle \le \langle \ell_i,p \rangle$ for all $i\in\LAs$
 and the coordinates of $p$ are positive.
\item[(c)] Coordinates of $v$ satisfy $\sum_{j=1}^M v(j)=0$.
\end{itemize}
For any $\epsilon \in (0,\alpha)$,
\begin{itemize}
\item[(d)] $p_1 = p + \epsilon v \in \Delta_M$
 and $p_2 = p - \epsilon v \in \Delta_M$,
\item[(e)] for any $i\in\LAs$, $i \ne i_1$,
 we have $\langle \ell_i-\ell_{i_1},p_1 \rangle \ge c \epsilon$,
\item[(f)] for any $i\in\LAs$, $i \ne i_2$,
 we have $\langle \ell_i-\ell_{i_2},p_2 \rangle \ge c \epsilon$.
\end{itemize}
\end{lemma}

\begin{proof}[Proof of Lemma~\ref{lem:epsilon-close-distributions}]
For any action $i\in\LAs$,
 consider the cell
$$
 C_i = \{ p\in\Delta_M ~:~
 \forall i'\in\LAs, \ \langle \ell_i,p \rangle \le \langle \ell_{i'},p \rangle \}
$$
 in the probability simplex $\Delta_M$.
The cell $C_i$ corresponds to the set of outcome distributions under which action $i$ is optimal.
Each cell is the intersection of some closed half-spaces and $\Delta_M$,
 and thus it is a compact convex polytope of dimension at most $M-1$.
Note that
\begin{equation}
\label{eq:cellscoverDeltaM}
 \bigcup_{i=1}^N C_i = \Delta_M.
\end{equation}
For $C\subseteq\Delta_M$,
 denote $\Int C$
 its interior in the topology induced by the hyperplane
 $\{x \in \R^M ~:~ \langle (1,\dots,1),x \rangle = 1\}$
 and $\rInt C$ its relative interior%
\footnote{\emph{Relative interior} of $C\subseteq\R^M$ is
 its interior in the topology induced by the smallest affine space containing it.}.
Let $\Lebesgue$ be the $(M-1)$-dimensional Lebesgue-measure.
It is easy to see that
 for any pair of cells $C_i$, $C_{i'}$, $C_{i'}\cap\Int C_i=\emptyset$,
 that is, 
 $\Lebesgue(C_i\cap C_{i'})=0$,
 and so 
\begin{equation}
\label{eq:intCi_unique}
 \Int C_i\subseteq C_i\setminus\bigcup_{i'\ne i} C_{i'}.
\end{equation}
Hence the cells form a cell-decomposition of the simplex.
Any two cells $C_i$ and $C_{i'}$ are separated by the hyperplane
 $f_{i,i'} = \{ x \in \R^M ~:~ \langle \ell_i,x \rangle = \langle \ell_{i'},x \rangle \}$.
Note that $C_i \cap C_{i'}\subset f_{i,i'}$.
The cells are characterized by the following lemma (which itself holds also with duplicate actions): 
\begin{lemma}\label{lem:celltypes}
Action $i$ is dominated
 $\Leftrightarrow C_i\subseteq\bigcup_{i':\ell_{i'}\ne \ell_i} C_{i'}$ $\Leftrightarrow$
 $\Int C_i=\emptyset$ $\Leftrightarrow$ $\Lebesgue(C_i)=0$,
 that is, $C_i$ is $(M-1)$-dimensional (has positive $\Lebesgue$-measure) if and only if there is $p\in C_i\setminus\bigcup_{i':\ell_{i'}\ne \ell_i} C_{i'}$.
Hence there is three kind of ``cells'':
\begin{enumerate}
\item $C_i=\emptyset$ (action $i$ is never optimal),
\item $C_i\ne\emptyset$ has dimension less than $M-1$, $\Int C_i=\emptyset$, $\Lebesgue(C_i)=0$, $C_i\subseteq\bigcup_{i':\ell_{i'}\ne \ell_i} C_{i'}$ (action $i$ is degenerate),
\item action $i$ is non-dominated, $C_i$ is $(M-1)$-dimensional, $\rInt C_i=\Int C_i\ne\emptyset$, $\Lebesgue(C_i)>0$, there is $p\in C_i\setminus\bigcup_{i':\ell_{i'}\ne \ell_i} C_{i'}$
.
\end{enumerate}
Moreover $\bigcup_{i\not\in\D} C_i=\Delta_M$ for the set $\D$ of dominated actions.
\end{lemma}
The proof is in the Appendix.

The non-triviality of the game ($K\ge 2$) means that there are at least two non-dominated actions of type 3 above.
In the cell decomposition,
 due to Lemma~\ref{lem:celltypes}, 
 there must exist two such $(M-1)$-dimensional cells $C_{i_1}$ and $C_{i_2}$
 corresponding to two non-dominated actions $i_1$,$i_2$,
 such that their intersection $C_{i_1} \cap C_{i_2}$ is an $(M-2)$-dimensional polytope.
Clearly, $\ell_{i_1}\ne\ell_{i_2}$, since otherwise the cells would coincide;
 thus part (a) is satisfied.

Moreover, $\rInt(C_{i_1} \cap C_{i_2})\subseteq\rInt\Delta_M$
 since otherwise $\Lebesgue(C_{i_1})$ or $\Lebesgue(C_{i_2})$ would be zero.
We can choose any $p\in\rInt(C_{i_1} \cap C_{i_2})$.
This choice of $p$ guarantees that $p\in f_{i_1,i_2}$,
 $\langle \ell_{i_1},p \rangle = \langle \ell_{i_2},p \rangle$
 , $p\in\rInt\Delta_M$, and part (b) is satisfied.
Since $C_{i_1} \cap C_{i_2}$ is $(M-2)$-dimensional,
 it also implies that
 there exists $\delta > 0$ such that
 the $\delta$-neighborhood $\{ q \in \R^M ~:~ \|p-q\| < \delta \}$ of $p$
 is contained in $\rInt(C_{i_1} \cup C_{i_2})$.

Since
 $p \in f_{i_1,i_2}$ therefore
 the hyperplane of vectors satisfying (c) does not coincide with $f_{i_1,i_2}$
 implying that
 we can choose $v\in\R^M\setminus\{0\}$ satisfying part (c)
, $\|v\|<\delta$, and $v\not\in f_{i_1,i_2}$.
We can assume
\begin{equation}
\label{eq:v-direction}
 \langle \ell_{i_2}-\ell_{i_1},v \rangle > 0
\end{equation}
 (otherwise we choose $\,-v$).
Since $p \pm v$ lie in the $\delta$-neighborhood of $p$,
 they lie in $\rInt(C_{i_1} \cup C_{i_2})$.
In particular, since
 $\langle \ell_{i_1},p+v \rangle < \langle \ell_{i_2},p+v \rangle$
 and $\langle \ell_{i_2},p-v \rangle < \langle \ell_{i_1},p-v \rangle$,
 $p+v\in\rInt C_{i_1}$ and $p-v\in\rInt C_{i_2}$.
Let
\begin{equation}
\label{eq:p1p2def}
 p_1 = p+\epsilon v \qquad\text{and}\qquad p_2 = p-\epsilon v \;.
\end{equation}
The convexity of $C_{i_1}$ and $C_{i_2}$ implies that for any $\epsilon \in (0,1]$,
 $p_1\in\rInt C_{i_1}$ and $p_2\in\rInt C_{i_2}$.
This, in particular, ensures that $p_1$,$p_2\in\Delta_M$ and part (d) holds.

To prove (e) define
 $\I = \{i\in\LAs:\ell_i$ is collinear with 
 $\ell_{i_1}$ and $\ell_{i_2}\}$.
We consider two cases:
As the first case fix action $i \in \I\setminus\{i_1\}$,
 that is,
 $\ell_i$ is an affine combination $\ell_i = a_i\ell_{i_1} + b_i \ell_{i_2}$ for some $a_i+b_i=1$.
Since $i_1$ and $i_2$ are non-dominated, this must be a convex combination with $a_i$,$b_i \ge 0$.
There is no duplicate action, thus $\ell_i \ne \ell_{i_1}$ implying $b_i\ne 0$.
Hence $b_i>0$, and from \eqref{eq:p1p2def} for any $\epsilon \ge 0$
$$
 \langle \ell_i - \ell_{i_1},p_1 \rangle
 = \langle b_i\ell_{i_2} - b_i\ell_{i_1},p + \epsilon v \rangle
 = \epsilon b_i \langle \ell_{i_2}-\ell_{i_1},v \rangle
 \ge c \epsilon
$$
provided that $0 < c \le \min_{i\in\I\setminus\{i_1\}} b_i \langle \ell_{i_2}-\ell_{i_1},v \rangle = c'$.
From \eqref{eq:v-direction} we know that
 $b_i \langle \ell_{i_2}-\ell_{i_1},v \rangle$ and so $c'$ are positive.

As the second case
 suppose $i \not\in \I$.
Then, the hyperplane $f_{i_1,i}$ does not coincide with $f_{i_1,i_2}$.
Since $p\in\rInt(C_{i_1} \cap C_{i_2})$,
 $p\in f_{i_1,i}$ would contradict to $f_{i_1,i} \cap \rInt C_{i_1} = \emptyset$ implied by \eqref{eq:intCi_unique}.
Thus $p \in C_{i_1} \setminus f_{i_1,i}$
 and therefore $\langle \ell_{i_1},p \rangle < \langle \ell_i,p \rangle$.
This means that if we choose $0<c \le
 \min(c',\frac12\min_{i\not\in\I}\langle \ell_i-\ell_{i_1},p \rangle)$
 (that is positive and depends only on $\bL$ and not on $T$) then
 for $\epsilon < \alpha =
 \min(1,c/\max_{i\not\in\I}|\langle \ell_i-\ell_{i_1},v \rangle|)$,
 from \eqref{eq:p1p2def}
 we have again
$$
 \langle \ell_i - \ell_{i_1},p_1 \rangle \ge
 2c + \epsilon \langle \ell_i-\ell_{i_1},v \rangle
 > c >
 c \epsilon \;.
$$

Part (f) is proved analogously to part (e),
 and by adjusting $\alpha$ and $c$ if necessary.
\end{proof}

We now continue with a technical lemma,
 which quantifies an upper bound
 on the Kullback-Leibler (KL) divergence (or relative entropy) between the two distributions from the previous lemma.
Recall that the KL divergence between two probability distributions $p$,$q \in \Delta_M$ is defined as
$$
 D(p~\|~q) = \sum_{j=1}^M p_j \ln \left(\frac{p_j}{q_j}\right) \;.
$$
%
\begin{lemma}[KL divergence of $\epsilon$-close distributions]
\label{lem:kl-divergence}
Let $p \in \Delta_M$ be a probability vector.
For any vector $\ep \in \R^M$ such that 
 both $p-\ep$ and $p+\ep$ lie in $\Delta_M$
 and $|\ep(j)|\le p(j)/2$ for all $j\in\NAs$,
 the KL divergence of $p-\ep$ and $p+\ep$ satisfies
$$
 D(p-\ep~\|~p+\ep) \le c \|\ep\|^2
$$
for some constant $c$ depending only on $p$.
\end{lemma}

\begin{proof}[Proof of Lemma~\ref{lem:kl-divergence}]
Since $p$, $p+\ep$, and $p-\ep$ are all probability vectors,
 notice that the coordinates of $\ep$ have to sum up to zero.
Also 
 if a coordinate of $p$ is zero then
 the corresponding coordinate of $\ep$ has to be zero as well.
As zero coordinates do not modify the KL divergence,
 we can assume without loss of generality that all coordinates of $p$ are positive.
By definition,
$$
 D(p-\ep~\|~p+\ep)
 = \sum_{j=1}^M (p(j)-\ep(j)) \ln \left( \frac{p(j) - \ep(j)}{p(j) + \ep(j)} \right) \;.
$$
We write the logarithmic factor as
$$
 \ln \left( \frac{p(j)-\ep(j)}{p(j)+\ep(j)} \right) =
 \ln \left(1-\frac{\ep(j)}{p(j)}\right) - \ln \left(1+\frac{\ep(j)}{p(j)}\right) \;.
$$
We use the second order Taylor expansion
 $\ln(1 \pm x) = \pm x - x^2/2 + O(|x|^3)$ around $0$
 to get that $\ln(1-x)-\ln(1+x) = -2x+r(x)
$,
 where $r(x)$ is a remainder upper bounded
 for all $|x| \le 1/2$ as $|r(x)|\le c'|x|^3$ with some universal constant $c' > 0$.%
\footnote{In fact, one can take $c' = 8\ln(3/e) \approx 0.79$.}
Substituting
\begin{align*}
 D(p-\ep~\|~p+\ep)
& = \sum_{j=1}^M (p(j)-\ep(j))
 \left[-2\frac{\ep(j)}{p(j)}+r\left(\frac{\ep(j)}{p(j)}\right)\right]\\
& = -2 \sum_{j=1}^M \ep(j) + 2 \sum_{j=1}^M \frac{\ep^2(j)}{p(j)}
 + \sum_{j=1}^M (p(j)-\ep(j)) \cdot r\left(\frac{\ep(j)}{p(j)}\right) \;.
\end{align*}
Here the first term is $0$.
Letting $\underline{p}=\min_{j\in\NAs} p(j)$,
 the second term is bounded by
 $2\sum_{j=1}^M \ep^2(j)/\underline{p} = (2/\underline{p}) \|\ep\|^2$,
 and the third term is bounded by
\begin{align*}
 \sum_{j=1}^M (p(j)-\ep(j)) \left|r\left(\frac{\ep(j)}{p(j)}\right)\right|
& \le 
 c' \sum_{j=1}^M (p(j)-\ep(j)) \frac{|\ep(j)|^3}{p^3(j)}
 = c' \sum_{j=1}^M \left( \frac{|\ep(j)|}{p(j)}-\frac{\ep(j)|\ep(j)|}{p^2(j)} \right) \frac{\ep^2(j)}{p(j)}\\
& \le c' \sum_{j=1}^M \left( \frac{|\ep(j)|}{p(j)}+\frac{|\ep(j)|^2}{p^2(j)} \right) \frac{\ep^2(j)}{p(j)}\\
& \le c' \sum_{j=1}^M \left( \frac12+\frac14 \right) \frac{\ep^2(j)}{\underline{p}}
 = \frac{3c'}{4\underline{p}} \|\ep\|^2 \;.
\end{align*}
Hence,
 $D(p-\ep~\|~p+\ep) \le \frac{8+3c'}{4\underline{p}} \|\ep\|^2 = c \|\ep\|^2$ for $c=\frac{8+3c'}{4\underline{p}}$.
\end{proof}

\begin{proof}[Proof of Theorem~\ref{theorem:sqrt_lower_bound_nontriv_games}]
The proof is similar as in~\citet{Auer-Cesa-Bianchi-Freund-Schapire-2002}.
When $M=1$, $G$ is always trivial, thus we assume that $M\ge 2$.
Without loss of generality we may assume that all the actions are \allrev.
Then, as in Section~\ref{section:characterization} for M=2,
 we can also assume that there are no duplicate actions,
 thus for any two actions $i$ and $i'$, $\ell_i\ne\ell_{i'}$.

Lemma~\ref{lem:epsilon-close-distributions} implies that
 there exist two actions $i_1$,$i_2$, $p\in\Delta_M$, $v\in\R^M$, and $c_1$,$\alpha > 0$
 satisfying conditions (a)--(f).
To avoid cumbersome indexing,
 by renaming the actions we can achieve that $i_1=1$ and $i_2=2$.
Let $p_1 = p + \epsilon v$ and $p_2 = p - \epsilon v$ for some $\epsilon \in (0,\alpha)$.
We determine the precise value of $\epsilon$ later.
By Lemma~\ref{lem:epsilon-close-distributions} (d), $p_1$,$p_2\in\Delta_M$.

Fix any randomized learning algorithm $A$ and time horizon $T$.
We use randomization replacing the outcomes
 by a sequence $J_1,J_2,\dots,J_T$ of random variables
 i.i.d.\ according to $p_k$, $k\in\{1,2\}$,
 and independently of the internal randomization of $A$.
Let
\begin{equation}
\label{eq:defNik}
 N_i^{(k)} = N_i^{(k)}(A,T) = \sum_{t=1}^T \Prn_k[I_t=i]\in[0,T]
\end{equation}
 be the expected number of times action $i$ is chosen by $A$ under $p_k$ up to time step $T$.
With subindex $k$,
 $\Pr_k$ and $\Exp_k$ denote probability and expectation
 given outcome model $k\in\{1,2\}$, respectively.
\begin{lemma}
\label{lem:supj_Nik}
For any partial-monitoring game with $N$ actions and $M$ outcomes, algorithm $A$
 and outcome distribution $p_k\in\Delta_M$
 such that action $k$ is optimal under $p_k$,
 we have
\begin{equation}
\label{eq:easy-lower-bound-regret1}
 \wcRegret_T(A,G)
 \ge \sum_{\substack{i\in\LAs \\ i\ne k}} N_i^{(k)} \langle \ell_i-\ell_k,p_k \rangle,
 \qquad k=1,2 \;.
\end{equation}
\end{lemma}
The proof is in the Appendix.

Parts (e) and (f) of Lemma~\ref{lem:epsilon-close-distributions} imply that
 $\langle \ell_k,p_k \rangle \le \langle \ell_i,p_k \rangle$
 for $k\in\{1,2\}$ and any $i\in\LAs$,
 hence $\wcRegret_T(A,G)$ can be bounded in terms of $N_i^{(k)}$
 using Lemma~\ref{lem:supj_Nik}.
They also imply that 
 for any $i\in\LAs$
 if $\ell_i\ne\ell_k$ then $\langle \ell_i-\ell_k,p_k \rangle \ge c_1 \epsilon$.
Therefore, we can continue lower bounding \eqref{eq:easy-lower-bound-regret1} as
\begin{equation}
\label{eq:easy-lower-bound-regret3}
 \sum_{\substack{i\in\LAs \\ i\ne k}} N_i^{(k)} \langle \ell_i-\ell_k,p_k \rangle
 \ge \sum_{\substack{i\in\LAs \\ i\ne k}} N_i^{(k)} c_1 \epsilon
 = c_1 \left(T-N_k^{(k)}\right) \epsilon \;.
\end{equation}
Collecting \eqref{eq:easy-lower-bound-regret1}
 and \eqref{eq:easy-lower-bound-regret3},
 we see that the worst-case regret of $A$ is lower bounded by
\begin{equation}
\label{eq:easy-lower-bound-regret4}
 \wcRegret_T(A,G) \ge c_1 \left(T-N_k^{(k)}\right) \epsilon
\end{equation}
for $k\in\{1,2\}$.
Averaging \eqref{eq:easy-lower-bound-regret4} over $k\in\{1,2\}$
 we get
\begin{equation}
\label{eq:supRge_average}
 \wcRegret_T(A,G) \ge c_1 \left( 2T-N_1^{(1)}-N_2^{(2)} \right)\epsilon/2 \;.
\end{equation}
We now focus on lower bounding $2T-N_1^{(1)}-N_2^{(2)}$.
We start by showing that
 $N_2^{(2)}$ is close to $N_2^{(1)}$.
The following lemma, which is the key lemma of both lower bound proofs,
 carries that out formally and states that
 the expected number of times an action is played by $A$
 does not change too much when we change the model,
 if the outcome distributions $p_1$ and $p_2$ are ``close'' in KL-divergence:
\begin{lemma}
\label{lem:N12Dp1p2gen}
For any partial-monitoring game with $N$ actions and $M$ outcomes, algorithm $A$,
 pair of outcome distributions $p_1$,$p_2\in\Delta_M$ and action $i$,
 we have
\begin{align*}
 N_i^{(2)}- N_i^{(1)} & \le T\sqrt{D(p_2~\|~p_1) \Nr^{(2)}/2}
& \text{ and } & &
 N_i^{(1)} -N_i^{(2)} & \le T\sqrt{D(p_1~\|~p_2) \Nr^{(1)}/2},
\end{align*}
where $\Nr^{(k)}=\sum_{t=1}^T \Prn_k[I_t\in\RR]=\sum_{i\in\RR} N_i^{(k)}$ under model $p_k$, $k=1$,$2$
 with $\RR$ being the set of revealing actions.%
\footnote{It seems from the proof that $\Nr^{(k)}$ could be slightly sharpened to $\Nr^{(k,T-1)}=\sum_{t=1}^{T-1} \Prn_k[I_t\in\RR]$.}
\end{lemma}
The proof is in the Appendix.

We use Lemma~\ref{lem:N12Dp1p2gen} for $i=2$
 and that $\Nr^{(2)}\le T$ to bound the difference $N_2^{(2)}-N_2^{(1)}$ as
\begin{equation}
\label{eq:easy-lower-bound-pull-ack2-difference}
 N_2^{(2)}-N_2^{(1)} \le T\sqrt{D(p_2~\|~p_1)T/2} = T^{3/2}\sqrt{D(p_2~\|~p_1)/2} \;.
\end{equation}
We upper bound $D(p_2~\|~p_1)$ using Lemma~\ref{lem:kl-divergence} with $\ep=\epsilon v$.
The lemma implies that
 $D(p_2~\|~p_1) \le c_2\epsilon^2$ for $\epsilon<\epsilon_0$
 with some $\epsilon_0$, $c_2>0$ which depend only on $v$ and $p$.
Putting this together with \eqref{eq:easy-lower-bound-pull-ack2-difference}
 we get
$$
 N_2^{(2)} < 
 N_2^{(1)} + c_3 \epsilon T^{3/2}
$$
where $c_3=\sqrt{c_2/2}$.
Together with $N_1^{(1)} + N_2^{(1)} \le T$ we get
$$
 2T - N_1^{(1)} - N_2^{(2)}
 > 2T - N_1^{(1)} - N_2^{(1)} - c_3 \epsilon T^{3/2}
 \ge T-c_3\epsilon T^{3/2} \;.
$$
Substituting into \eqref{eq:supRge_average}
 and choosing $\epsilon = 1/(2c_3 T^{1/2})$
 gives
 the desired lower bound
$$
 \wcRegret_T(A,G) > \frac{c_1}{8c_3} \sqrt{T}
$$
provided that our choice of $\epsilon$ ensures that
 $\epsilon < \min(
\alpha,\epsilon_0
)=:\epsilon_1$
 that depends only on $\bL$.
This condition is satisfied for all $T > T_0 =
 1/(2c_3\epsilon_1)^2$.
Since $c_1$, $c_3$,
 and $\epsilon_1$ depend only on $\bL$, for such $T$, $\Regret_T(G) \ge
 \tfrac{c_1}{8c_3} \sqrt{T}$.

The non-triviality of the game
 implies that
 Lemma~\ref{lem:0minimaxregret} d) does not hold, so neither does b), that is,
 $\Regret_T(G)>0$ for $T\ge 1$.
Thus choosing
$$
 c
 = \min\left( \min_{1\le T\le T_0} \frac{\Regret_T(G)}
 {\sqrt{T}},\frac{c_1}{8c_3}
 \right),
$$
$c>0$
 and for any $T$, $\Regret_T(G) \ge
 c \sqrt{T}$.
\end{proof}

\noindent\textbf{Remark}\quad
Theorem~\ref{theorem:sqrt_lower_bound_nontriv_games} also holds if $M=\infty$.
Namely,
 since the proof of c)$\ra$d) of Lemma~\ref{lem:0minimaxregret} remains obviously valid,
 the non-triviality of the game ($K\ge 2$) excludes that c) holds,
 and thus for each $i\in\LAs$ there is $j_i\in\{1,2,\dots\}$ such that
 $\ell_{i,j_i}$ is not minimal in the $j_i^{\rm th}$ column of $\bL$.
Then take the minor of $\bL$ consisting of its (at most $N$) columns corresponding to $\OO=\{j_1,\dots,j_N\}$.
For the corresponding finite game $G_\OO$ (that does not depend on $A$),
 Lemma~\ref{lem:0minimaxregret} c) still does not hold,
 thus nor d) does, and $G_\OO$ is also non-trivial.
Hence Theorem~\ref{theorem:sqrt_lower_bound_nontriv_games} implies that%
\footnote{The same reasoning can be used to show that
 we could assume without loss of generality $M\le N$ in the proof of Theorem~\ref{theorem:sqrt_lower_bound_nontriv_games}.}
$$
 \Regret_T(G) = \inf_A\sup_{j_{1:T}\in\{1,2,\dots\}^T} \Regret_T(A,G)
 \ge \inf_A\sup_{j_{1:T}\in\OO^T} \Regret_T(A,G)
 = \Regret_T(G_\OO) = \Omega\left(\sqrt{T}\right) \;.
$$

\section{Lower Bound for Hard Games}
\label{section:lower-bound-hard-games}
In this section,
 we present an $\Omega(T^{2/3})$ lower bound for the expected regret of any two-outcome game
 in the case when the separation condition does not hold.
\begin{theorem}[Lower bound for hard games]
\label{theorem:lower-hard}
If $M=2$ and $G=(\bL,\bH)$ satisfies the non-de\-gen\-er\-a\-cy condition
 and the separation condition does \textbf{not} hold
 then there exists a constant $C>0$ such that for any $T\ge 1$
 the minimax expected regret $\Regret_T(G)\ge CT^{2/3}$.
\end{theorem}

\comment{ 
We need the following result.
\begin{lemma}
\label{lem:bern}
Let $\alpha \in (0,1)$.
For any $\epsilon \in [0,\min\{\alpha, 1-\alpha\})$
$$
 D(\alpha-\epsilon~\|~\alpha+\epsilon) = O(\epsilon^2) \qquad \text{as \ $\ep \to 0$}
$$
where the constant and the interval in the $O(\cdot)$ notation depend only on $\alpha$.
\end{lemma}

\begin{proof}[Proof of Lemma~\ref{lem:bern}]
Follows from Lemma~\ref{lem:kl-divergence} with $M=2$, $p=(\alpha,1-\alpha)^\top$, and $\ep=(\epsilon,-\epsilon)^\top$.
\end{proof}
} 

\begin{proof}[Proof of Theorem~\ref{theorem:lower-hard}]
We follow the lower bound proof for the label efficient prediction
 from~\citet{Cesa-Bianchi-Lugosi-Stoltz-2006} with a few changes.
The most important change, as we will see, is the choice of the models we randomize over.

As the first step, the following lemma shows that \nonrev\ degenerate actions do not influence the minimax regret of a game.
\begin{lemma}\label{lem:degenerate}
Let $G$ be a non-degenerate game with two outcomes.
Let $G'$ be the game we get by removing the degenerate \nonrev\ actions from $G$.
Then $\Regret_T(G)=\Regret_T(G')$.
\end{lemma}
The proof of this lemma can be found in the Appendix.

By the non-degeneracy condition and Lemma~\ref{lem:degenerate},
 we can assume without loss of generality that $G$ does not have degenerate actions.
We can also assume without loss of generality that
 actions $1$ and $2$ are the two consecutive non-dominated \nonrev\ actions.
It follows by scaling and a reduction
 similar to the one we used in Section~\ref{section:algorithms} that
 we can further assume $(\ell_{1,1},\ell_{1,2})=(0,\alpha)$,
 $(\ell_{2,1},\ell_{2,2})=(1-\alpha,0)$ with some $\alpha\in(0,1)$.
Using the non-de\-gen\-er\-a\-cy condition and
 that actions 1 and 2 are consecutive non-dominated actions,
 we get that for all $i \ge 3$,
 there exists some $\lambda_i\in\R$ depending only on $\bL$ such that
\begin{align}
\label{cond:lower}
\begin{split}
\ell_{i,1} & > \lambda_i\ell_{1,1}+(1-\lambda_i)\ell_{2,1} = (1-\lambda_i)(1-\alpha)\;,\\
\ell_{i,2} & > \lambda_i\ell_{1,2}+(1-\lambda_i)\ell_{2,2} = \lambda_i \alpha \;.
\end{split}
\end{align}
Let $\lamin=\min_{i \ge 3}\lambda_i$, $\lamax=\max_{i \ge 3}\lambda_i$,
 and $\lambda^*=\lamax-\lamin$.

We define two models for generating outcomes from $\{1,2\}$.
In model $1$, the outcome distribution is $p_1(1)=\alpha+\epsilon$, $p_1(2)=1-p_1(1)$,
 whereas in model $2$, $p_2(1)=\alpha-\epsilon$, $p_2(2)=1-p_2(1)$
 with $0<\epsilon \le\min(\alpha,1-\alpha)/2$ to be chosen later.
We use randomization replacing the outcomes
 by a sequence $J_1,J_2,\dots,J_T$ of random variables
 i.i.d.\ according to $p_k$, $k\in\{1,2\}$,
 and independently of the internal randomization of $A$.
Let
$
 N_i^{(k)} 
$
 be the expected number of times action $i$ is chosen by $A$ under $p_k$ up to time step $T$,
 as in \eqref{eq:defNik}.
With subindex $k$,
 $\Pr_k$ and $\Exp_k$ denote probability and expectation
 given outcome model $k\in\{1,2\}$, respectively.
Finally, let $N_{\ge 3}^{(k)}=\sum_{i \ge 3} N_i^{(k)}$.
Note that, if $\epsilon<\epsilon_0$ with some $\epsilon_0$ depending only on $\bL$
 then only actions $1$ and $2$ can be optimal for these models.
Namely, action $k$ is optimal under $p_k$,
 hence $\wcRegret_T(A,G)$ can be bounded in terms of $N_i^{(k)}$
 using Lemma~\ref{lem:supj_Nik}:
\begin{equation}
\label{eq:hard-lower-bound-regret1}
 \wcRegret_T(A,G)
 \ge \sum_{\substack{i\in\LAs \\ i\ne k}} N_i^{(k)} \langle \ell_i-\ell_k,p_k \rangle
 = \sum_{i=3}^N N_i^{(k)} \langle \ell_i-\ell_k,p_k \rangle
 + N_{3-k}^{(k)} \langle \ell_{3-k}-\ell_k,p_k \rangle
\end{equation}
for $k=1$,$2$.
Now, by~\eqref{cond:lower}, there exists $\tau>0$ depending only on $\bL$ such that for all $i \ge 3$,
$
\ell_{i,1} \ge (1-\lambda_i)(1-\alpha) + \tau
$
and
$
\ell_{i,2} \ge \alpha\lambda_i + \tau
$.
These bounds and simple algebra give that
\begin{align*}
 \langle \ell_i-\ell_1,p_1 \rangle
&= (\ell_{i,1}-\ell_{1,1})(\alpha+\epsilon) + (\ell_{i,2}-\ell_{1,2})(1-\alpha-\epsilon)\\
&\ge ((1-\lambda_i)(1-\alpha)+\tau)(\alpha+\epsilon)
 + (\alpha\lambda_i+\tau-\alpha)(1-\alpha-\epsilon)\\
&= (1-\lambda_i)\epsilon +\tau\\
&\ge (1-\lamax)\epsilon +\tau =: f_1
\end{align*}
 and
$$
 \langle \ell_2-\ell_1,p_1 \rangle
 = (1-\alpha)(\alpha+\epsilon) - \alpha(1-\alpha-\epsilon)
 = \epsilon \;.
$$
Analogously, we get
$$
 \langle \ell_i-\ell_2,p_2 \rangle
 \ge \lamin\epsilon +\tau =: f_2
 \qquad\text{and}\qquad
 \langle \ell_1-\ell_2,p_2 \rangle
 = \epsilon \;.
$$
Note that if $\epsilon<\tau/\max(|1-\lamax|,|\lamin|)$
 then both $f_1$ and $f_2$ are positive.
Substituting these into \eqref{eq:hard-lower-bound-regret1} gives
\begin{equation}\label{eq:12lower}
 \wcRegret_T(A,G) \ge f_k N_{\ge 3}^{(k)} + \epsilon N_{3-k}^{(k)} \;.
\end{equation}
The following lemma is an application of Lemma~\ref{lem:N12Dp1p2gen} and~\ref{lem:kl-divergence}:
\begin{lemma}
\label{lem:N12}
There exists a constant $c > 0$ (depending on $\alpha$ only) such that
\begin{align*}
 N_2^{(1)} & \ge N_2^{(2)}-cT\epsilon\sqrt{N_{\ge 3}^{(2)}}
& \text{ and } & &
 N_1^{(2)} & \ge N_1^{(1)}-cT\epsilon\sqrt{N_{\ge 3}^{(1)}} \;.
\end{align*}
\end{lemma}

\begin{proof}
We only prove the first inequality, the other one is symmetric.
Using Lemma~\ref{lem:N12Dp1p2gen} with $M=2$, $i=2$ and the fact that actions $1$ and $2$ are \nonrev,
 we have
$$
 N_2^{(2)}-N_2^{(1)} \le T\sqrt{D(p_2~\|~p_1) N_{\ge 3}^{(2)}/2} \;.
$$
Lemma~\ref{lem:kl-divergence} with $M=2$, $p=(\alpha,1-\alpha)^\top$, and $\ep=(\epsilon,-\epsilon)^\top$
 gives $D(p_2~\|~p_1) \le \hat{c}\epsilon^2$,
 where $\hat{c}$ depends only on $\alpha$.
Rearranging and substituting $c=\sqrt{\hat{c}/2}$ yields the first statement of the lemma.
\end{proof}

Let $l=\arg\min_{k\in\{1,2\}} N_{\ge 3}^{(k)}$.
Now, for $k\neq l$
 we can lower bound the regret 
 using Lemma~\ref{lem:N12} for \eqref{eq:12lower}:
\begin{equation}\label{eq:hard-lower-bound-regret2}
 \wcRegret_T(A,G)
 \ge f_k N_{\ge 3}^{(k)} + \epsilon\left(N_{3-k}^{(l)}-cT\epsilon\sqrt{N_{\ge 3}^{(l)}}\right)
 \ge f_k N_{\ge 3}^{(l)} + \epsilon\left(N_{3-k}^{(l)}-cT\epsilon\sqrt{N_{\ge 3}^{(l)}}\right),
\end{equation}
 as $f_k>0$.
For $k=l$ we do this
 subtracting $cT\epsilon^2\sqrt{N_{\ge 3}^{(l)}}\ge 0$
 from the right-hand side of~\eqref{eq:12lower} leading to the same lower bound,
 hence \eqref{eq:hard-lower-bound-regret2} holds for $k=1$,$2$.
Finally,
 averaging \eqref{eq:hard-lower-bound-regret2} over $k\in\{1,2\}$
 we have the bound
\begin{align*}
 \frac{f_1+f_2}{2} N_{\ge 3}^{(l)} +
 \epsilon\left(\frac{N_2^{(l)}+N_1^{(l)}}{2}-cT\epsilon\sqrt{N_{\ge 3}^{(l)}}\right)
&= \left( \frac{(1-\lamax+\lamin)\epsilon}{2} + \tau \right) N_{\ge 3}^{(l)}
 + \epsilon\left(\frac{T-N_{\ge 3}^{(l)}}{2}\right)-cT\epsilon^2\sqrt{N_{\ge 3}^{(l)}}\\
&= \left(\tau-\frac{\lambda^*\epsilon}{2}\right) N_{\ge 3}^{(l)}
 + \frac{\epsilon T}{2} - cT\epsilon^2\sqrt{N_{\ge 3}^{(l)}} \;.
\end{align*}
Choosing $\epsilon=c_2 T^{-1/3}\,(\le c_2)$ with $c_2>0$
 gives
\begin{align*}
 \wcRegret_T(A,G)
&\ge \left(\tau-\frac{\lambda^* c_2 T^{-1/3}}{2}\right) N_{\ge 3}^{(l)}
 + \frac{c_2 T^{2/3}}{2} - c c_2^2 T^{1/3}\sqrt{N_{\ge 3}^{(l)}}\\
&\ge \left(\tau-\frac{\lambda^* c_2}{2}\right) N_{\ge 3}^{(l)}
 + \frac{c_2 T^{2/3}}{2} - c c_2^2 T^{1/3}\sqrt{N_{\ge 3}^{(l)}}\\
&= \left( \left(\tau-\frac{\lambda^* c_2}{2}\right) x^2
 + \frac{c_2}{2} - c c_2^2 x \right) T^{2/3}
 = q(x) T^{2/3},
\end{align*}
where $x=T^{-1/3}\sqrt{N_{\ge 3}^{(l)}}$ and $q(x)$ can be written and lower bounded as
$$
 q(x) 
 = \left(\tau-\frac{\lambda^* c_2}{2}\right) \left(x-\frac{c c_2^2}{2\tau-\lambda^* c_2}\right)^2
 + \frac{c_2}{2} - \frac{c^2 c_2^4}{4\tau-2\lambda^* c_2}
 \ge \frac{c_2}{2} \left(1-\frac{c^2 c_2}{2\tau-\lambda^* c_2}\right)
$$
independently of $x$ whenever $\lambda^* c_2<2\tau$ and $c_2\le 1$.
Now it is easy to see that if $c_2=\min(\tau/(c^2+\lambda^*),1)$ then these hold,
 moreover,
$q(x) \ge 
 c_2/4>0$ 
 giving
 the desired lower bound
$$
 \wcRegret_T(A,G) \ge \frac{c_2}{4} T^{2/3}
$$
provided that our choice of $\epsilon$ ensures that
 $\epsilon < \min(
\alpha/2,(1-\alpha)/2,\epsilon_0,\tau/|1-\lamax|,\tau/|\lamin|
)=:\epsilon_1$
 that depends only on $\bL$.
This condition is satisfied for all $T > T_0 =
 (c_2/\epsilon_1)^3$.
Since $c_2$
 and $\epsilon_1$ depend only on $\bL$, for such $T$, $\Regret_T(G) \ge
 \tfrac{c_2}{4} T^{2/3}$.

If the separation condition does not hold then the game is clearly non-trivial which,
 using Lemma~\ref{lem:0minimaxregret} b) and d) as in the proof of Theorem~\ref{theorem:sqrt_lower_bound_nontriv_games},
 implies that
 $\Regret_T(G)>0$ for $T\ge 1$.
Thus choosing
$$
 C
 = \min\left( \min_{1\le T\le T_0} \frac{\Regret_T(G)}
 {T^{2/3}},\frac{c_2}{4}
 \right),
$$
$C>0$
 and for any $T$, $\Regret_T(G) \ge
 C T^{2/3}$.
\end{proof}

\section{Discussion}
\label{section:discussion}

In this paper we classified non-degenerate partial-monitoring games with two outcomes based on their minimax regret.
An immediate question is how the classification extends to degenerate games.
Unfortunately, the degeneracy condition is needed in both the upper and lower bound proofs. We do not even know if all degenerate games fall into one of the four categories or there are some games with minimax regret of $\widetilde\Theta(T^\alpha)$ for some $\alpha\in(1/2,2/3)$.
Nonetheless, we conjecture that,
 if the revealing degenerate actions are included in the chain of non-dominated actions,
 the classification theorem holds without any change.

The most important open question is
 whether our results generalize to games with more outcomes.
A simple observation is that, given a finite partial-monitoring game, if we
 restrict \nature's choices to any two outcomes, the resulting game's hardness
 serves as a lower bound on the minimax regret of the original game.
This gives us a sufficient condition that a game has $\Omega(T^{2/3})$ minimax regret.
We believe that the $\Omega(T^{2/3})$ lower bound can also be generalized to
 situations where two ``$\epsilon$-close'' outcome distributions are not
 distinguishable by playing only their respective optimal actions.
Generalizing the upper bound result seems more challenging.
The algorithm \AppleTree{} heavily exploits the two-dimensional structure of the losses and, as of yet,
 in general we do not know how to construct an algorithm
 that achieves $\widetilde O(\sqrt{T})$ regret on partial-monitoring games with more than two outcomes.

It is also important to note that our upper bound result heavily exploits the assumption that the opponent is oblivious. Our results do not extend to games with non-oblivious opponents, to the best of our knowledge.

\appendix
\section{}
\label{section:appendix}

\begin{proof}[Proof of Lemma~\ref{lem:0minimaxregret}]
a)$\ra$b) is obvious.

b)$\ra$c)
For any $A$,
\begin{align*}
 \wcRegret_T(A,G)
&\ge \sup_{j\in\NAs, J_1=\cdots=J_T=j} \Exp\left[ \sum_{t=1}^T \ell_{I_t,J_t} - \min_{i\in\LAs} \sum_{t=1}^T \ell_{i,J_t}\right]\\
&= \sup_{j\in\NAs} \Exp\left[\sum_{t=1}^T \ell_{I_t,j} - T\min_{i\in\LAs} \ell_{i,j}\right]\\
&\ge \sup_{j\in\NAs} \left(\Exp\left[\ell_{I_1,j}\right] - \min_{i\in\LAs} \ell_{i,j}\right)
 = f(A) \;.
\end{align*}
b) leads to
\[
 0 = \Regret_T(G) = \inf_A \wcRegret_T(A,G) \ge \inf_A f(A) \;.
\]
Observe that $f(A)$ depends on $A$ through
 only the distribution of $I_1$ on $\LAs$ denoted by $q=q(A)$ now, 
 that is, $f(A)=f'(q)$ for proper $f'$.
This dependence is continuous on the compact domain of $q$,
 hence the infimum can be replaced by minimum.
Thus $\min_q f'(q)\le 0$, that is, there exists a $q$ such that for all $j\in\NAs$,
 $\Exp\left[\ell_{I_1,j}\right] = \min_{i\in\LAs} \ell_{i,j}$.
This implies that the support of $q$ contains only actions
 whose loss is not larger than the loss of any other action
 irrespectively of the choice of Nature's action.
(Such an action is obviously non-dominated as shown by any $p\in\Delta_M$ supported on all outcomes.)

c)$\ra$d)
Action $i$ in c) is non-dominated,
 and any other action with loss vector distinct from $\ell_i$ is dominated
 (by $i$ and any action with loss vector $\ell_i$).

d)$\ra$a)
For any action $i\in\LAs$,
 as in the proof of Lemma~\ref{lem:epsilon-close-distributions},
 consider the compact convex cell
$
 C_i 
$
 in 
 $\Delta_M$
.
By Lemma~\ref{lem:celltypes} $\bigcup_{i\not\in\D} C_i=\Delta_M$.
This and d) imply that there is an $i$ with $C_i=\Delta_M$, that is,
 $i$ is optimal for any outcome.
So the algorithm that always plays $i$ has zero regret for all outcome sequences and $T$.
\end{proof}

\begin{proof}[Proof of Theorem~\ref{theorem:characterization} Case~\eqref{eq:char4}]
We know that $K\ge 2$ and $G$ has no revealing action.
Then for any $A$,
\begin{align*}
 \wcRegret_T(A,G)
&\ge \sup_{j\in\NAs, J_1=\cdots=J_T=j} \Exp\left[ \sum_{t=1}^T \ell_{I_t,J_t} - \min_{i\in\LAs} \sum_{t=1}^T \ell_{i,J_t}\right]\\
&\ge \frac1{M}\sum_{j=1}^M \Exp\left[\sum_{t=1}^T \ell_{I_t,j} - T\min_{i\in\LAs} \ell_{i,j}\right]\\
&= \frac1{M}\sum_{t=1}^T \Exp\left[\sum_{j=1}^M \ell_{I_t,j}\right]
 - \frac{T}{M}\sum_{j=1}^M \min_{i\in\LAs} \ell_{i,j} \;.
\end{align*}
Here $I_t$ is a random variable usually depending on $J_{1:T-1}$, that is, on $j$ through the outcomes.
However, since $G$ has no revealing action,
 now the distribution of $I_t$ is independent of $j$, thus
 $\Exp[\sum_{j=1}^M \ell_{I_t,j}] \ge \min_{i\in\LAs} \sum_{j=1}^M \ell_{i,j}$
 for each $t$,
 and we have
$$
 \wcRegret_T(A,G)
 \ge T \underbrace{\frac1{M} \left[ \min_{i\in\LAs} \sum_{j=1}^M \ell_{i,j}
 - \sum_{j=1}^M \min_{i\in\LAs} \ell_{i,j} \right]}_c
 = cT\;,
$$
where $c>0$ if $K\ge 2$
 (because $c\ge 0$, and $c=0$ would imply Lemma~\ref{lem:0minimaxregret} c), thus also d)).
Since $c$ depends only on $\bL$, $\Regret_T(G)\ge cT=\Theta(T)$.
\end{proof}

\begin{proof}[Proof of Lemma~\ref{lem:celltypes}]
By Definition~\ref{def:action_prop}, action $i$ is dominated
 if and only if $C_i\subseteq\bigcup_{i':\ell_{i'}\ne \ell_i} C_{i'}$.

$C_i\subseteq\bigcup_{i':\ell_{i'}\ne \ell_i} C_{i'} \ra \Int C_i=\emptyset$:
Since $\ell_{i'}\ne \ell_i\ra i\ne i$, follows from \eqref{eq:intCi_unique}.

$\Int C_i=\emptyset \ra \Lebesgue(C_i)=0$:
Follows from convexity of $C_i$.

$\Lebesgue(C_i)=0 \ra C_i\subseteq\bigcup_{i':\ell_{i'}\ne \ell_i} C_{i'}$:
 indirect: if $p\in C_i$ is in the complementer of $\bigcup_{i':\ell_{i'}\ne \ell_i} C_{i'}$,
 that is open in $\Delta_M$,
 then there is a neighborhood $S$ of $p$ in $\Delta_M$ disjoint from $\bigcup_{i':\ell_{i'}\ne \ell_i} C_{i'}$.
Thus $S\subseteq \bigcup_{i':\ell_{i'}=\ell_i} C_{i'}=C_i$ due to \eqref{eq:cellscoverDeltaM},
 and $\Lebesgue(C_i)\ge\Lebesgue(S)>0$, contradiction.

Since $\Lebesgue(\bigcup_{i\in\D} C_i)\le\sum_{i\in\D}\Lebesgue(C_i)=0$,
 thus from \eqref{eq:cellscoverDeltaM} $\Lebesgue(\bigcup_{i\not\in\D} C_i)\ge\Lebesgue(\Delta_M)$,
 and $\Lebesgue(\Delta_M\setminus\bigcup_{i\not\in\D} C_i)=0$.
The latest set is open in $\Delta_M$, so it must be empty,
 that is, $\bigcup_{i\not\in\D} C_i=\Delta_M$.
\end{proof}

\begin{proof}[Proof of Lemma~\ref{lem:supj_Nik}]
Clearly, the worst-case expected regret of $A$ is at least its average regret:
$$
 \wcRegret_T(A,G) = \sup_{j_{1:T}\in\NAs^T} \Regret_T(A,G)
 \ge \Exp_k[\Regret_T(A,G)] 
 = \Exp_k \left[ \sum_{t=1}^T \ell_{I_t,J_t} - \min_{i\in\LAs} \sum_{t=1}^T \ell_{i,J_t} \right]\;,
$$
where 
 the expectation on the right-hand side
 is taken with respect to both the random choices of the outcomes
 and the internal randomization of $A$.
We lower bound the right-hand side switching expectation and minimum to get
{\allowdisplaybreaks
\begin{align}
\label{eq:easy-lower-bound-regret2}
\notag
 \Exp_k \left[ \sum_{t=1}^T \ell_{I_t,J_t} - \min_{i\in\LAs} \sum_{t=1}^T \ell_{i,J_t} \right]
& \ge \sum_{t=1}^T \Exp_k\ell_{I_t,J_t} - \min_{i\in\LAs} \sum_{t=1}^T \Exp_k\ell_{i,J_t}\\
\notag
& = \sum_{t=1}^T \sum_{i=1}^N \Exp_k\left[ \one{I_t=i}\ell_{i,J_t} \right]
 - \min_{i\in\LAs} \sum_{t=1}^T \langle \ell_i,p_k \rangle\\
\notag
& = \sum_{t=1}^T \sum_{i=1}^N \Exp_k\one{I_t=i} \Exp_k\ell_{i,J_t}
 - T \min_{i\in\LAs} \langle \ell_i,p_k \rangle\\
\notag
& \mbox{(by the independence of $I_t$ and $J_t$)}\\
\notag
& = \sum_{i=1}^N \langle \ell_i,p_k \rangle \sum_{t=1}^T \Prn_k[I_t=i]
 - T \min_{i\in\LAs} \langle \ell_i,p_k \rangle\\
& = \sum_{i=1}^N N_i^{(k)} \langle \ell_i,p_k \rangle - T \langle \ell_k,p_k \rangle\\
\notag 
&= \sum_{\substack{i\in\LAs \\ i\ne k}} N_i^{(k)} \langle \ell_i-\ell_k,p_k \rangle \;.
\end{align}}
\eqref{eq:easy-lower-bound-regret2} follows from the fact that action $k$ is optimal under $p_k$.
Clearly the term $i=k$ can be omitted in the last equality.
\end{proof}

\begin{proof}[Proof of Lemma~\ref{lem:N12Dp1p2gen}]
We only prove the first inequality, the other one is symmetric.
Assume first that $A$ is deterministic,
 that is, $I_t:\Sigma^{t-1} \to \LAs$,
 and so $I_t(h_{1:t-1})$ denotes the choice of the algorithm at time step $t$,
 given that the (random) history of observations of length $t-1$,
 $H_{1:t-1}=(H_1,\dots,H_{t-1})$ takes $h_{1:t-1}=(h_1,\dots,h_{t-1}) \in 
 \Sigma^{t-1}$.
(Note that this is a slightly different history definition than $\H_{1:t-1}$ defined in Section~\ref{section:algorithms}, as
$H_{1:t-1}$ does not include the actions since their choices are determined by the feedback anyway.
In general, $\H_{1:t-1}$ is equivalent to $H_{1:t-1} \cup (I_1,...,I_{t-1})$.
Nevertheless, if it is assumed that the feedback symbol sets of actions are disjoint
 then $H_{1:t-1}$ and $\H_{1:t-1}$ are equivalent.) 
We denote by $p_k^*$ the joint distribution of $H_{1:T-1}$  over $\Sigma^{T-1}$ associated with $p_k$.
(For games with only \allrev\ actions, assuming $h_{i,j}=j$ in $\bH$,
 $p_k^*$ is the product distribution over the outcome sequences,
 that is, formally, $p_k^*(j_{1:T-1}) = \prod_{t=1}^{T-1} p_k(j_t)$.)
We can bound the difference $N_2^{(2)}-N_2^{(1)}$ as
\begin{align}
\label{eq:lower-bound-pull-ack2-difference}
\notag N_i^{(2)} - N_i^{(1)}
&= \sum_{t=1}^T \left( \Prn_2[I_t=i] - \Prn_1[I_t=i] \right)\\
\notag &= \sum_{h_{1:T-1}\in\Sigma^{T-1}} \sum_{t=1}^T \left(
 \one{I_t(h_{1:t-1})=i} p^*_2(h_{1:T-1}) - \one{I_t(h_{1:t-1})=i} p^*_1(h_{1:T-1})
 \right)\\
\notag &= \sum_{h_{1:T-1}\in\Sigma^{T-1}} \left(p^*_2(h_{1:T-1})-p^*_1(h_{1:T-1})\right) \cdot
 \sum_{t=1}^T \one{I_t(h_{1:t-1})=i}\\
&\le T \sum_{\substack{h_{1:T-1}\in\Sigma^{T-1} \\ p^*_2(h_{1:T-1}) \ge p^*_1(h_{1:T-1})}}
 \left(p^*_2(h_{1:T-1})-p^*_1(h_{1:T-1})\right)\\
\notag &= \frac{T}{2} \left\|p^*_2-p^*_1\right\|_1\\
\notag &\le T\sqrt{D(p^*_2~\|~p^*_1)/2}\;,
\end{align}
 where the last step is
 an application of Pinsker's inequality \citep[Lemma~12.6.1]{Cover-Thomas-2006} to distributions $p_1^*$ and $p_2^*$.
%
Using the chain rule 
 for KL divergence \citep[Theorem~2.5.3]{Cover-Thomas-2006} we can write (with somewhat sloppy notation)
$$
 D(p^*_2~\|~p^*_1)
 = \sum_{t=1}^{T-1} D\left(p^*_2(h_t~|~h_{1:t-1})~\|~p^*_1(h_t~|~h_{1:t-1})\right)\;,
$$
where the $t^{\rm th}$ conditional KL divergence term is
\begin{align}
\label{eq:condKLdiv}
 \sum_{h_{1:t-1}\in\Sigma^{t-1}} \Prn_2(H_{1:t-1}=h_{1:t-1}) \sum_{h_t\in\Sigma}
 \Prn_2(H_t=h_t~|~H_{1:t-1}=h_{1:t-1}) \ln\frac{\Prn_2(H_t=h_t~|~H_{1:t-1}=h_{1:t-1})}{\Prn_1(H_t=h_t~|~H_{1:t-1}=h_{1:t-1})} \;.
\end{align}
Decompose this sum for the case $I_t(h_{1:t-1})\not\in\RR$ and $I_t(h_{1:t-1})\in\RR$.
In the first case, we play a none-revealing action, thus our observation
 $H_t=h_{I_t(h_{1:t-1}),J_t}=h_{I_t(h_{1:t-1}),1}$ is a deterministic constant in both models $1$ and $2$,
 thus both $\Prn_1(\cdot~|~H_{1:t-1}=h_{1:t-1})$ and $\Prn_2(\cdot~|~H_{1:t-1}=h_{1:t-1})$ are degenerate
 and the KL divergence factor is 0.
Otherwise, playing a revealing action,
 $H_t$$=h_{I_t(h_{1:t-1}),J_t}$ is the same deterministic function of $J_t$
 (which is independent of $H_{1:t-1}$) in both models $1$ and $2$,
 and so the inner sum in \eqref{eq:condKLdiv} is
\begin{align}
\label{eq:KLdiv_given_hist}
 \sum_{h_t\in\Sigma} \Prn_2[h_{I_t(h_{1:t-1}),J_t}=h_t]
 \ln\frac{\Prn_2[h_{I_t(h_{1:t-1}),J_t}=h_t]}{\Prn_1[h_{I_t(h_{1:t-1}),J_t}=h_t]} \;.
\end{align}
Since $\Prn_k[h_{I_t(h_{1:t-1}),J_t}=h_t] = \sum_{j_t\in\NAs:h_{I_t(h_{1:t-1}),j_t}=h_t} p_k(j_t)$ ($k=1$,$2$),
 using the log sum inequality \citep[Theorem 2.7.1]{Cover-Thomas-2006}),
 \eqref{eq:KLdiv_given_hist} is upper bounded by
$$
 \sum_{h_t\in\Sigma}\sum_{j_t\in\NAs:h_{I_t(h_{1:t-1}),j_t}=h_t} p_2(j_t) \ln\frac{p_2(j_t)}{p_1(j_t)}
 = \sum_{j_t\in\NAs} p_2(j_t) \ln\frac{p_2(j_t)}{p_1(j_t)}
 = D(p_2~\|~p_1) \;.
$$
Hence, $D(p^*_2~\|~p^*_1)$ is upper bounded by
$$
 \sum_{t=1}^{T-1}\sum_{\substack{h_{1:t-1}\in\Sigma^{t-1} \\ I_t(h_{1:t-1})\in\RR}}
 \Prn_2(H_{1:t-1}=h_{1:t-1}) D(p_2~\|~p_1)
 = D(p_2~\|~p_1) \sum_{t=1}^{T-1}\sum_{i\in\RR} \Prn_2[I_t=i]
 = D(p_2~\|~p_1) \Nr^{(2,T-1)}
\;,
$$
where $\Nr^{(k,T-1)}=\sum_{t=1}^{T-1} \Prn_k[I_t\in\RR]$.
This together with \eqref{eq:lower-bound-pull-ack2-difference} 
 gives $N_i^{(2)} - N_i^{(1)} \le T\sqrt{D(p_2~\|~p_1) \Nr^{(2,T-1)}/2}$.

If $A$ is random and its internal random ``bits'' are represented by a random value $Z$
 (which is independent of $J_1$,$J_2$,\dots),
 then $N_i^{(k)} 
 =\Exp\left[\tilde{N}_i^{(k)}(Z)\right]$
 for $\tilde{N}_i^{(k)}(Z)=\sum_{t=1}^T\Prn_k[I_t=i|Z]$.
Also let $\tNr^{(k,T-1)}(Z)=\sum_{t=1}^{T-1} \Prn_k[I_t\in\RR|Z]$.
The proof above implies that for any fixed $z\in{\rm Range}(Z)$,
$$
 \tilde{N}_i^{(2)}(z) - \tilde{N}_i^{(1)}(z)
 \le T\sqrt{D(p_2~\|~p_1) \tNr^{(2,T-1)}(z)/2}\;,
$$
and thus, using also Jensen's inequality,
\begin{align*}
 N_i^{(2)} - N_i^{(1)}
&= \Exp\left[ \tilde{N}_i^{(2)}(Z) - \tilde{N}_i^{(1)}(Z) \right]\\
&\le \Exp\left[ T\sqrt{D(p_2~\|~p_1) \tNr^{(2,T-1)}(Z)/2} \right]\\
&\le T\sqrt{D(p_2~\|~p_1) \Exp\left[\tNr^{(2,T-1)}(Z)\right]/2}
 = T\sqrt{D(p_2~\|~p_1) \Nr^{(2,T-1)}/2}\;,
\end{align*}
that is clearly upper bounded by $T\sqrt{D(p_2~\|~p_1) \Nr^{(2)}/2}$ yielding the statement of the lemma.
\end{proof}

\begin{proof}[Proof of Lemma~\ref{lem:degenerate}]
We prove the lemma by showing that
 for every algorithm $A$ on game $G$ there exists an algorithm $A'$ on $G'$
 such that for any outcome sequence, $\Regret_T(A',G')\le\Regret_T(A,G)$ and vice versa.
Recall that the minimax regret of a game is
  \begin{align*}
    \Regret_T(G) &= \inf_A \ \sup_{J_{1:T} \in \NAs^T} \ \Regret_T(A,G)\;,\\
    \intertext{where}
    \Regret_T(A,G)&= \Exp\left[\sum_{t=1}^T \ell_{I_t,J_t} - \min_{i\in\LAs} \sum_{t=1}^T \ell_{i,J_t}\right]\;,
  \end{align*}
First we observe that the term $\Exp[\min_{i\in\LAs} \sum_{t=1}^T \ell_{i,J_t}]$ does not change by removing degenerate actions.
Indeed, by the definition of degenerate action,
 if the minimum is given by a degenerate action then there exists a non-degenerate action with the same cumulative loss.
It follows that we only have to deal with the term $\Exp[\sum_{t=1}^T \ell_{I_t,J_t}]$.
\begin{enumerate}
 \item Let $A'$ be an algorithm on $G'$.
We define the algorithm $A$ on $G$ by choosing the same actions as $A'$ at every time step.
Since the action set of $G$ is a superset of that of $G'$,
 this construction results in a well defined algorithm on $G$,
 and trivially has the same expected loss as $A'$.
 \item Let $A$ be an algorithm on $G$.
From the definition of degenerate actions,
 we know that for every degenerate action $i$, there are two possibilities:
 \begin{enumerate}
  \item There exists a non-degenerate action $i_1$
 such that $\ell_i$ is component-wise lower bounded by $\ell_{i_1}$.
  \item There are two non-degenerate actions $i_1$ and $i_2$
 such that $\ell_i$ is a convex combination of $\ell_{i_1}$ and $\ell_{i_2}$,
 that is, $\ell_i=\alpha_i\ell_{i_1}+(1-\alpha_i)\ell_{i_2}$ for some $\alpha_i\in(0,1)$.
 \end{enumerate}
 \begin{figure}
 \centering
   \includegraphics{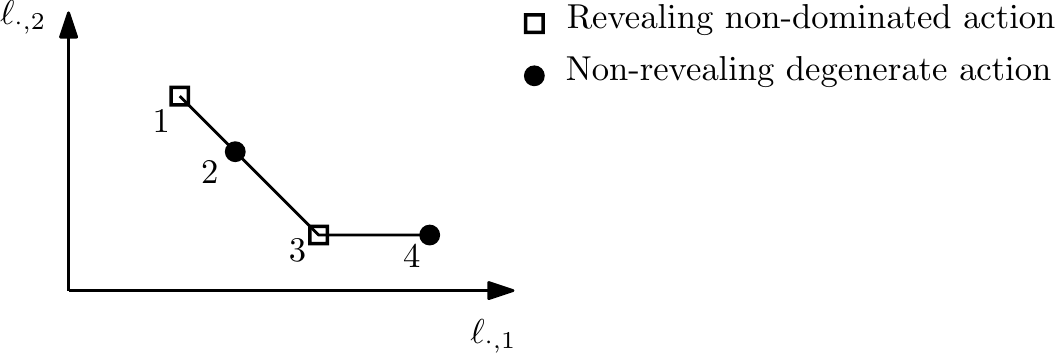}
   \caption{Degenerate \nonrev\ actions on the chain. The loss vector of action 2 is a convex combination of that of action 1 and 3. On the other hand, the loss vector of action 4 is component-wise lower bounded by that of action 3.}
   \label{fig:deglemma}
 \end{figure}
An illustration of these cases can be found in Figure~\ref{fig:deglemma}.
We construct $A'$ the following way.
At every time step $t$, if $I_t^{A}$ (the action that algorithm $A$ would take) is non-degenerate then let $I_t^{A'}=I_t^{A}$.
If $I_t^A=i$ is a degenerate action of the first kind, let $I_t^{A'}$ be $i_1$.
If $I_t^A=i$ is a degenerate action of the second kind
 then let $I_t^{A'}$ be $i_1$ with probability $\alpha_i$ and $i_2$ with probability $1-\alpha_i$.
Recall that $G$ is non-degenerate, so $i$ has to be a \nonrev\ action.
However, $i_1$ and/or $i_2$ might be revealing ones.
To handle this, $A'$ is defined to map the observation sequence,
 before using it as the argument of $I_t$, replacing the feedbacks corresponding to degenerate action $i$ by $h_{i,1}=h_{i,2}$.
That is, intuitively, $A'$ ``pretends'' that the feedbacks at such time steps are irrelevant.
It is clear that the expected loss of $A'$ in every time step is less than or equal to the expected loss of $A$,
 concluding the proof.
\end{enumerate}
\end{proof}

\subsection*{Proof of Theorem~\ref{theorem:sqrt_lower_bound_nontriv_games} for adversarial nature}

For the proof, we start with a lemma,
 which ensures the existence of a pair $i_1$,$i_2$ of actions
 and an outcome distribution $p$ with $M$ atoms such
 that both $i_1$ and $i_2$ are optimal under $p$.
\begin{lemma}
\label{lem:2opt_for_pos_distr}
Let $G=(\bL,\bH)$ be any finite non-trivial game with $N$ actions and $M\ge 2$ outcomes.
Then there exists $p\in\Delta_M$
 satisfying both of the following properties:
\begin{itemize}
\item[(a)] All coordinates of $p$ are positive.
\item[(b)] There exist actions $i_1$,$i_2\in\LAs$ such that
 $\ell_{i_1}\neq\ell_{i_2}$ and 
 for all $i\in\LAs$,
$$
 \langle \ell_{i_1},p \rangle = \langle \ell_{i_2},p \rangle
 \le \langle \ell_i,p \rangle \;.
$$
\end{itemize}
\end{lemma}

\begin{proof}[Proof of Lemma~\ref{lem:2opt_for_pos_distr}]
Note that distributions $p$ with positive coordinates form the interior of $\Delta_M$ ($\Int\Delta_M$).
For any action $i\in\LAs$,
 as in the proof of Lemma~\ref{lem:epsilon-close-distributions},
 consider the compact convex cell
$
 C_i 
$
 in 
 $\Delta_M$,
 whose union is $\Delta_M$ (see
 \eqref{eq:cellscoverDeltaM}).
Let $p_1$ be any point in the interior of $\Delta_M$.
By \eqref{eq:cellscoverDeltaM}, there is a cell $C_{i_1}$ containing $p_1$.
If $C_{i_1}=\Delta_M$ held 
 then 
 action $i_1$ would
 satisfy Lemma~\ref{lem:0minimaxregret} c),
 thus also d), and the game would be trivial.
So there must be a point, say $p_2$, in $\Delta_M\setminus C_{i_1}$.
The intersection of the closed segment $\overline{p_1 p_2}$ and $C_{i_1}$ is closed and convex,
 thus it is a closed subsegment $\overline{p_1 p}$ for some $p\in C_{i_1}$ ($p\ne p_2$).
$p_1\in\Int\Delta_M$ and the convexity of $\Delta_M$ imply $p\in\Int\Delta_M$.
Since the open segment $\overline{p p_2}$ has to be covered by
 $\bigcup_{i':C_{i'}\ne C_{i_1}} C_{i'}$, that is a closed set,
 $p\in\bigcup_{i:C_{i'}\ne C_{i_1}} C_{i'}$ must also hold,
 that is, $p\in C_{i_2}$ for some $C_{i_2}\ne C_{i_1}$ (requiring $\ell_{i_1}\ne \ell_{i_2}$).
Hence $p$ satisfies both (a) and (b).
\end{proof}

\begin{proof}[Proof of Theorem~\ref{theorem:sqrt_lower_bound_nontriv_games}]
When $M=1$, $G$ is always trivial, thus we assume that $M\ge 2$.
Without loss of generality we may assume that all the actions are \allrev.

Let $p\in\Delta_M$ be a distribution of the outcomes
 that satisfies conditions (a) and (b) of Lemma~\ref{lem:2opt_for_pos_distr}.
By renaming actions we can assume without loss of generality that
 $\ell_1 \neq \ell_2$ and actions $1$ and $2$ are optimal under $p$,
 that is,
\begin{equation}
\label{eq:12optforp}
 \langle \ell_1,p \rangle = \langle \ell_2,p \rangle
 \le \langle \ell_i,p \rangle
\end{equation}
for any $i\in\LAs$.

Fix any learning algorithm $A$.
We use randomization replacing the outcomes
 by a sequence $J_1,J_2,\dots,J_T$ of random variables i.i.d.\ according to $p$,
 and independent of the internal randomization of $A$.
Clearly, as in the proof of Lemma~\ref{lem:supj_Nik},
 the worst-case expected regret of $A$ is at least its average regret:
\begin{equation}
\label{eq:supRge_bound}
 \wcRegret_T(A,G)
 \ge \Exp[\Regret_T(A,G)] 
 = \Exp \left[ \sum_{t=1}^T \ell_{I_t,J_t} - \min_{i\in\LAs} \sum_{t=1}^T \ell_{i,J_t} \right]
 = \Exp \left[ \sum_{t=1}^T \Exp[\ell_{I_t,J_t}~|~I_t] - \min_{i\in\LAs} \sum_{t=1}^T \ell_{i,J_t} \right] \;.
\end{equation}
Here, 
 in the last two expressions,
 the expectation is with respect to
 both the internal randomization of $A$ and the random choice of $J_1, J_2, \dots, J_T$.
Now, since $J_t$ is independent of $I_t$,
 we see that $\Exp[\ell_{I_t,J_t} ~|~ I_t] = \langle \ell_{I_t},p \rangle$.
By \eqref{eq:12optforp}, 
 we have $\langle \ell_{I_t},p \rangle \ge \langle \ell_1,p \rangle = \langle \ell_2,p \rangle$.
Therefore (upper bounding also the minimum),
\begin{align}
\label{eq:Nto2actions}
 \sum_{t=1}^T \Exp \left[ \ell_{I_t,J_t}~|~I_t \right] - \min_{i\in\LAs} \sum_{t=1}^T \ell_{i,J_t}
& = \sum_{t=1}^T \langle \ell_{I_t},p \rangle - \min_{i\in\LAs} \sum_{t=1}^T \ell_{i,J_t}\notag\\
& \ge \sum_{t=1}^T \langle \ell_1,p \rangle - \min_{i=1,2} \sum_{t=1}^T \ell_{i,J_t}\\
& = \max_{i=1,2} \sum_{t=1}^T \left( \langle \ell_1,p \rangle - \ell_{i,J_t} \right)\notag
 \;.
\end{align}
Using the identity $\max\{a,b\} = \frac12(a+b + |a-b|)$,
 the latest expression is
\begin{align*}
& \frac12 \left[ \sum_{t=1}^T \left( \langle \ell_1,p \rangle-\ell_{1,J_t} \right)
 + \sum_{t=1}^T \left( \langle \ell_1,p \rangle-\ell_{2,J_t} \right)
 + \left| \sum_{t=1}^T \left( \langle \ell_1,p \rangle-\ell_{1,J_t} \right)
 - \sum_{t=1}^T \left( \langle \ell_1,p \rangle-\ell_{2,J_t} \right) \right| \right]\\
& = \frac12 \sum_{t=1}^T \left( \langle \ell_1,p \rangle-\ell_{1,J_t} + \langle \ell_2,p \rangle-\ell_{2,J_t} \right)
 + \frac12 \left| \sum_{t=1}^T \left( \ell_{2,J_t}-\ell_{1,J_t} \right) \right|
,
\end{align*}
where \eqref{eq:12optforp} was used in the first term.
The expectation of the first term vanishes since $\Exp[\ell_{i,J_t}]=\langle \ell_i,p \rangle$. 
Let $X_t = 
 \ell_{2,J_t}-\ell_{1,J_t}$.
We see that $X_1, X_2, \dots, X_T$ are i.i.d.\ random variables with mean $\Exp[X_t] = 0$.
Therefore,
\begin{equation}
\label{eq:genKhinchine}
 \Exp\left[ \max_{i=1,2} \sum_{t=1}^T \left( \langle \ell_1,p \rangle - \ell_{i,J_t} \right) \right]
 = \frac12 \Exp \left| \sum_{t=1}^T X_t \right|
 \ge c \sqrt{T},
\end{equation}
where the last inequality follows from Theorem~\ref{theorem:generalized-khinchine-inequality} stated below
 and the constant $c$ depends only on $\ell_1, \ell_2$, and $p$.
For the theorem to yield $c > 0$,
 it is important to note that
 the distribution of $X_t$ has finite support and
 with positive probability $X_t \neq 0$ since $\ell_1 \neq \ell_2$ and all coordinates of $p$ are positive.
Hence, both $\Exp[X_t^2]$ and $\Exp[X_t^4]$ are finite and positive.

Now, putting together \eqref{eq:supRge_bound}, \eqref{eq:Nto2actions}, and \eqref{eq:genKhinchine}
 gives the desired lower bound $\wcRegret_T(A,G) \ge c \sqrt{T}$.
Since $c$ depends only on $\bL$, also $\Regret_T(G) \ge c \sqrt{T}$.
\end{proof}

The following theorem is a variant of Khinchine's inequality~(see e.g.~\cite[Lemma~A.9]{Cesa-Bianchi-Lugosi-2006}) 
 for asymmetric random variables.
The idea of the proof is the same as there 
 and originally comes from Littlewood~\cite{Littlewood-1930}.
%
\begin{theorem}[Khinchine's inequality for asymmetric random variables]
\label{theorem:generalized-khinchine-inequality}
Let $X_1, X_2, \dots, X_T$ be i.i.d.\ random variables with mean $\Exp[X_t] = 0$,
 finite variance $\Exp[X_t^2] = \Var(X_t) = \sigma^2$,
 and finite fourth moment $\Exp[X_t^4] = \mu_4$.
Then,
$$
 \Exp \left| \sum_{t=1}^T X_t \right|
 \ge \frac{\sigma^3}{\sqrt{3\mu_4}} \sqrt{T} \;.
$$
\end{theorem}

\begin{proof}
\cite[Lemma~A.4]{DeGyLu96} implies that
 for any random variable $Z$ with finite fourth moment
$$
\Exp |Z| \ge \frac{\left( \Exp[Z^2] \right)^{3/2}}{ \left( \Exp[Z^4] \right)^{1/2}} \;.
$$
Applying this inequality to $Z = \sum_{t=1}^T X_t$ we get
$$
\Exp \left| \sum_{t=1}^T X_t \right|
 \ge \frac{ T^{3/2} \sigma^3}{ T \sqrt{3\mu_4} }
 = \frac{\sigma^3}{\sqrt{3\mu_4}} \sqrt{T},
$$
that follows from
$$
\Exp[Z^2] = \Exp \left[ \left( \sum_{t=1}^T X_t \right)^2 \right]
= \sum_{t=1}^T \Exp[X_t^2]
= T \sigma^2
$$
and
$$
\Exp[Z^4] = \Exp \left[ \left( \sum_{t=1}^T X_t \right)^4 \right]
= \sum_{t=1}^T \Exp[X_t^4] + 6 \!\!\!\! \sum_{1 \le s < t \le T} \!\!\!\! \Exp[X_s^2] \Exp[X_t^2]
= T \mu_4 + 3T(T-1)\sigma^4
\le 3T^2 \mu_4,
$$
where we have used the independence of $X_t$'s and $\Exp[X_t]=0$
 which ensure that mixed terms $\Exp[X_tX_s]$, $\Exp[X_tX_s^3]$, etc. vanish.
We also used that 
 $\sigma^4 = \Exp[X_t^2]^2 \le \Exp[X_t^4] = \mu_4$.
\end{proof}

\bibliographystyle{unsrtnat}
\bibliography{biblio}	

\end{document}